\documentclass[10pt,english,journal]{IEEEtran}

\usepackage{amsmath,amssymb,amsfonts,mathtools}
\usepackage{amsthm}
\usepackage{bm}
\usepackage{booktabs}
\usepackage{algorithm}
\usepackage[noend]{algpseudocode}
\usepackage{graphicx}
\usepackage{microtype}
\usepackage[hidelinks]{hyperref}
\usepackage[nameinlink]{cleveref}
\usepackage{braket}
\crefname{table}{Table}{Tables}
\usepackage{url}
\usepackage{balance}
\usepackage{cite}
\usepackage{enumitem}
\usepackage{siunitx}
\usepackage{verbatim}
\usepackage{caption}
\usepackage{mdframed}
\usepackage[utf8]{inputenc}
\DeclareUnicodeCharacter{2009}{\,}   
\DeclareUnicodeCharacter{2013}{--}   

\newcommand{\R}{\mathbb{R}}
\newcommand{\E}{\mathbb{E}}

\newcommand{\norm}[1]{\left\lVert #1 \right\rVert}
\newcommand{\abs}[1]{\left\lvert #1 \right\rvert}
\newcommand{\ip}[2]{\left\langle #1,#2 \right\rangle}
\DeclareMathOperator{\diag}{diag}
\DeclareMathOperator{\Tr}{Tr}
\DeclareMathOperator{\Var}{Var}

\newcommand{\bigO}{\mathcal{O}}
\newcommand{\Cov}{\mathrm{Cov}}

\theoremstyle{plain}

\theoremstyle{remark}

\usepackage{tabularx}   
\usepackage{array}      
\allowdisplaybreaks 
\newcommand{\PP}{\mathbb{P}}
\newcommand{\cF}{\mathcal{F}}

\newcommand{\SSM}{\mathrm{SSM}}

\usepackage{hyphenat}

\usepackage{bbm}

\usepackage{nicefrac}

\usepackage{float}

\theoremstyle{definition}

\usepackage{adjustbox}
\usepackage{comment}

\usepackage[table]{xcolor}

\makeatletter
\renewcommand{\SSM}{\ifmmode\mathrm{SSM}\else\textsc{SSM}\fi}
\makeatother

\definecolor{ALGpre}{HTML}{CFE8FF}   
\definecolor{ALGssm}{HTML}{FFE3C2}   
\definecolor{ALGtt}{HTML}{DCEBFF}    
\definecolor{ALGtrain}{HTML}{FFD6D6} 
\definecolor{ALGinfer}{HTML}{D9F7D9} 

\definecolor{ALGpreLine}{HTML}{2B7BBB}
\definecolor{ALGssmLine}{HTML}{E07B18}
\definecolor{ALGttLine}{HTML}{2C55D4}
\definecolor{ALGtrainLine}{HTML}{C93131}
\definecolor{ALGinferLine}{HTML}{1F8F3A}

\title{LiQSS: Post-Transformer Linear Quantum-Inspired State-Space Tensor Networks for Real-Time 6G}

\author{
Farhad~Rezazadeh,~\IEEEmembership{Member,~IEEE},
Hatim~Chergui,~\IEEEmembership{Senior~Member,~IEEE},
Amir~Ashtari~Gargari,~\IEEEmembership{Member,~IEEE},
Mehdi Bennis,
Houbing Song,
Lingjia~Liu,~and~Merouane~Debbah,~\IEEEmembership{Fellow,~IEEE}

\IEEEcompsocitemizethanks{\IEEEcompsocthanksitem F. Rezazadeh is with the BrainOmega and the Technical University of Catalonia (UPC), 08028 Barcelona, Spain (e-mail: farhad.rezazadeh@upc.edu).}
\IEEEcompsocitemizethanks{\IEEEcompsocthanksitem H. Chergui is with i2CAT Foundation, 08034 Barcelona, Spain (e-mail: hatim.chergui@i2cat.net).}
\IEEEcompsocitemizethanks{\IEEEcompsocthanksitem A. shtari Gargari is with the Centre Tecnologic de Telecomunicacions de Catalunya
(CTTC/CERCA), 08860 Castelldefels, Spain (e-mail: aashtari@cttc.es).}
\IEEEcompsocitemizethanks{\IEEEcompsocthanksitem M. Bennis is with the University of Oulu, 90570 Oulu, Finland (e-mail: mehdi.bennis@oulu.fi).}
\IEEEcompsocitemizethanks{\IEEEcompsocthanksitem H. Song is with the University of Maryland, Baltimore County (UMBC), 21250 Baltimore, USA (e-mail: h.song@ieee.org).}
\IEEEcompsocitemizethanks{\IEEEcompsocthanksitem L. Liu is with the Virginia Tech, 24061 Blacksburg, USA (e-mail: ljliu@vt.edu).}
\IEEEcompsocitemizethanks{\IEEEcompsocthanksitem M. Debbah is with the Khalifa University of Science and Technology, 127788 Abu Dhabi, UAE (e-mail: merouane.debbah@ku.ac.ae).}

}
\begin{document}
\IEEEaftertitletext{\vspace{-1.5\baselineskip}} 
\maketitle

\begin{abstract}
Proactive and agentic control in Sixth-Generation (6G) Open Radio Access Networks (O-RAN) requires control-grade prediction under stringent Near-Real-Time (Near-RT) latency and computational constraints. While Transformer-based models are effective for sequence modeling, their quadratic complexity limits scalability in Near-RT RAN Intelligent Controller (RIC) analytics. This paper investigates a post-Transformer design paradigm for efficient radio telemetry forecasting. We propose a quantum-inspired many-body state-space tensor network that replaces self-attention with stable structured state-space dynamics kernels, enabling linear-time sequence modeling. Tensor-network factorizations in the form of Tensor Train (TT) / Matrix Product State (MPS) representations are employed to reduce parameterization and data movement in both input projections and prediction heads, while lightweight channel gating and mixing layers capture non-stationary cross-Key Performance Indicator (KPI) dependencies. The proposed model is instantiated as an agentic perceive--predict xApp and evaluated on a bespoke O-RAN KPI time-series dataset comprising 59{,}441 sliding windows across 13 KPIs, using Reference Signal Received Power (RSRP) forecasting as a representative use case. Our proposed Linear Quantum-Inspired State-Space (LiQSS)\footnote{To support reproducibility, the source code is publicly available for non-commercial use at \url{https://github.com/frezazadeh/quantum-state-space}.} model is $10.8\times$--$15.8\times$ smaller and approximately $1.4\times$ faster than prior structured state-space baselines. Relative to Transformer-based models, LiQSS achieves up to a $155\times$ reduction in parameter count and up to $2.74\times$ faster inference, without sacrificing forecasting accuracy.
\end{abstract}


\section{Introduction}

\IEEEPARstart{6}{G} mobile networks are expected to natively support intelligent, adaptive, and autonomous control across heterogeneous radio access technologies, extreme densification, and ultra-low latency services~\cite{wang2025ultralowlatency}. To this end, the O-RAN architecture introduces openness, disaggregation, and programmability as first-class design principles~\cite{agarwal2025openran6g}, enabling Artificial Intelligence (AI) and Machine Learning (ML) to be deployed as control and optimization functions in the form of xApps and rApps. In particular, the Near-RT RIC, operating at time scales from tens of milliseconds to seconds, is envisioned as a key enabler of proactive and agentic radio resource management in 6G networks.

A fundamental requirement for agentic control in the Near-RT RIC is \emph{control-grade prediction}; the ability to forecast radio telemetry and KPIs accurately and with bounded latency under tight computational and memory constraints. Accurate short-horizon forecasting of KPIs such as  RSRP, Signal-to-Interference-plus-Noise Ratio (SINR), and traffic load enables proactive scheduling, mobility management, interference mitigation, and energy optimization. However, Near-RT deployment imposes strict limits on inference latency, model footprint, and data movement, particularly when xApps are co-located with virtualized RAN components and must operate at scale across large numbers of cells and users.

Transformer-based architectures have recently emerged as a dominant paradigm for sequence modeling and time-series forecasting~\cite{zhou2021informer, zhou2022fedformer, woo2022etsformer, nie2023patchtst, zhang2023crossformer, peng2023rwkv, sun2023retentive, ansari2024chronos} due to their expressive self-attention mechanism and strong empirical performance across natural language processing, vision, and time-series analysis. Consequently, Transformers and their variants have been explored for radio analytics and network forecasting tasks. Nevertheless, the quadratic time and memory complexity of self-attention with respect to sequence length poses a significant barrier to their adoption in Near-RT RIC environments. This limitation is further exacerbated by the high dimensionality, multivariate nature, and non-stationarity of radio telemetry streams, making naive scaling of Transformer models impractical for real-time 6G control loops.

Motivated by these challenges, our recent research~\cite{rez2025rivaling, rezazadeh2025agentic} has explored \emph{post-Transformer} sequence modeling paradigms that replace explicit attention mechanisms with structured state-space dynamics. Linear State-Space Models (SSMs)~\cite{smith2023simplified}, particularly those derived from continuous-time systems and orthogonal polynomial projections, have demonstrated the ability to model long-range temporal dependencies with linear computational complexity. Among these, architectures based on High-order Polynomial Projection Operators (HiPPO) and their Legendre variants~\cite{gu2020hippo} enable stable and expressive representations of continuous-time signals while admitting efficient implementations via causal convolutions.

\begin{table*}[t!]
\centering
\scriptsize
\setlength{\tabcolsep}{8pt}
\caption{Overview of Representative Transformer- and SSM-Based Models for Time-Series.}
\label{tab:related_work_overview}
\resizebox{\textwidth}{!}{%
\begin{tabular}{p{2.6cm} p{3.0cm} p{3.4cm} p{3.6cm} p{3.8cm}}
\toprule
\textbf{Model} &
\textbf{Architecture} &
\textbf{Key Techniques} &
\textbf{Target Problems} &
\textbf{Limitations / Technical Gaps} \\
\midrule

Informer~\cite{zhou2021informer} &
Transformer (efficient attention; encoder--decoder) &
ProbSparse self-attention; attention distilling; generative-style decoder &
Long sequence time-series forecasting (LSTF) &
Still attention-based; long-context compute/memory remains nontrivial \\\\

FEDformer~\cite{zhou2022fedformer} &
Transformer with decomposition + frequency-domain modeling &
Seasonal-trend decomposition; Fourier / wavelet-enhanced blocks &
Long-term series forecasting (seasonality/trend + long horizons) &
Relies on decomposition and frequency priors; sensitive to mismatch with data characteristics \\\\

ETSformer~\cite{woo2022etsformer} &
Transformer-style forecasting model with smoothing components &
Exponential Smoothing Attention (ESA) and Frequency Attention (FA) &
Time-series forecasting with interpretability &
Smoothing-structured bias may limit flexibility for non-ETS dynamics; still Transformer-family complexity \\\\

PatchTST~\cite{nie2023patchtst} &
Transformer (patch-based; channel-independent design) &
Temporal patching; channel independence (weight sharing across channels) &
Long-term multivariate forecasting; representation learning / pretraining &
Patch hyperparameter sensitivity; channel-independence can miss cross-variable coupling \\\\

Crossformer~\cite{zhang2023crossformer} &
Transformer (cross-time + cross-dimension attention; hierarchical encoder--decoder) &
Dimension-Segment-Wise (DSW) embedding; Two-Stage Attention &
Multivariate forecasting with cross-variable dependencies &
Attention-centric and architecture-heavy; compute scales with cross-time/cross-dimension modeling \\\\

RWKV~\cite{peng2023rwkv} &
Transformer--RNN hybrid / linear-attention recurrence &
Time-mixing + channel-mixing blocks; linear-attention formulation &
General sequence modeling (incl.\ time series) &
General-purpose sequence model; limited time-series-specific forecasting validation in the reference \\\\

RetNet~\cite{sun2023retentive} &
Retention network (sequence model with retention mechanism) &
Retention with parallel/recurrent/chunkwise computation &
General sequence modeling (LLM-focused) &
LLM/sequence-model focus; limited time-series forecasting-specific validation and inductive bias discussion \\\\

Chronos~\cite{ansari2024chronos} &
Pretrained probabilistic time-series model via LM adaptation &
Scaling+quantization tokenization; pretrained Transformer LMs (multiple sizes) &
General time-series forecasting (zero-shot / pretrained) &
Model size and tokenization/decoding overhead can be costly; larger variants constrain deployment \\\\

\midrule
MS$^{3}$M~\cite{rez2025rivaling} &
Structured SSM forecaster (strictly causal) &
Multi-scale HiPPO--LegS kernels; Tustin discretization; SE gating; compact GLU mixing &
KPI forecasting (one-step; RSRP highlighted) &
Scope emphasizes strict causality and next-step prediction; longer-horizon/transfer needs separate study \\\\

WM--MS$^{3}$M~\cite{rezazadeh2025agentic} &
Agentic world model built on structured SSM mixtures &
Action-conditioned generative state-space; dual decoders; MPC/CEM planning &
What-if forecasting + short-horizon decision making &
World-model + planning stack adds complexity; performance depends on learned dynamics and action constraints \\
\bottomrule
\end{tabular}
}
\end{table*}

In parallel, tensor network methods~\cite{Dborin2022MPSPretraining} originating from quantum many-body physics~\cite{GlisicLorenzo2024QuantumNeuro6G7GSurvey, Tasaki2020QuantumManyBody} have gained attention as a principled approach to representing high-dimensional functions with reduced parameterization and controlled expressivity. The TT and MPS decompositions~\cite{
Dborin2022MPSPretraining, Bradley2020MLST}, in particular, provide low-rank factorizations that significantly reduce memory footprint and data movement, while preserving the ability to model complex correlations. Despite their theoretical appeal, the integration of tensor networks with state-space sequence models for real-time network intelligence remains largely unexplored.

In this paper, we propose a \emph{quantum-inspired linear state-space tensor network} architecture for efficient radio telemetry forecasting in 6G O-RAN systems. The proposed model replaces self-attention with structured, stable state-space dynamics based on stable HiPPO-LegS kernels, enabling linear-time sequence modeling. We employ tensor network factorizations in both input embeddings and prediction heads to reduce model size and improve computational efficiency. We introduce lightweight channel gating and mixing layers to capture non-stationary cross-KPI interactions without incurring the overhead of full attention mechanisms. Experimental results demonstrate that the proposed approach achieves accuracy competitive with Transformer-based baselines, while significantly reducing inference latency and model footprint. These findings support the viability of post-Transformer, state-space, and tensor-network-based models as enabling technologies for resource-efficient real-time intelligence in future 6G O-RAN deployments.

\subsection{Related Work}
\label{sec:related_work}

\subsubsection{Transformer-Based Time-Series Models}

Transformers have been widely adopted for time-series forecasting due to their ability to model long-range dependencies via self-attention.
Early adaptations such as Informer~\cite{zhou2021informer} introduced sparse attention mechanisms to mitigate the quadratic complexity of vanilla Transformers.
Subsequent studies explored frequency-domain decomposition and inductive biases, including FEDformer~\cite{zhou2022fedformer} and ETSformer~\cite{woo2022etsformer}, which explicitly model trend and seasonal components. More recent architectures have focused on structural modifications of the attention mechanism.
PatchTST~\cite{nie2023patchtst} segments time-series into patches to reduce effective sequence length,
while Crossformer~\cite{zhang2023crossformer} captures cross-dimensional dependencies by alternating temporal and feature-wise attention.
iTransformer~\cite{liu2024itransformer} inverts the role of temporal and feature dimensions, enabling efficient modeling of multivariate dependencies. Beyond classical Transformers, alternative attention approximations and recurrent-attention hybrids have been proposed.
Performers~\cite{choromanski2021rethinking} replaces softmax attention with kernelized random feature maps,
RWKV~\cite{peng2023rwkv} blends recurrent computation with attention-style parameterization,
and RetNet~\cite{sun2023retentive} introduces exponential decay-based retention to reduce memory overhead.
More recently, foundation-style models such as Chronos~\cite{ansari2024chronos} adapt large pretrained language models (T5/GPT) to time-series forecasting. Despite their strong empirical performance, Transformer-based models suffer from several limitations in Near-RT O-RAN settings:
(i) quadratic or near-quadratic scaling with sequence length,
(ii) large parameter footprints and memory traffic,
and (iii) weak alignment with the physical and dynamical structure of radio telemetry.
These limitations significantly hinder deployment within Near-RT RICs, where inference latency and resource efficiency are first-class constraints.

\subsubsection{Structured State-Space Models for Sequence Modeling}

Structured SSMs have recently emerged as a compelling post-Transformer alternative.
These models replace attention with linear dynamical systems that admit efficient convolutional or recurrent implementations.
A key development in this direction is the use of HiPPO-based~\cite{gu2022hippo} operators, which provide stable and expressive representations of continuous-time signals via orthogonal polynomial projections. Building on this foundation, several SSM architectures have demonstrated strong performance on long-sequence tasks with linear complexity.
In the context of radio intelligence and agentic control, our previous studies, MS$^{3}$M~\cite{rez2025rivaling} introduced multi-scale structured state-space mixtures for Near-RT O-RAN analytics,
while WM--MS$^{3}$M~\cite{rezazadeh2025agentic} extended this framework toward agentic world modeling and near-real-time reasoning.
These studies demonstrated that SSM-based predictors can outperform attention-based models under strict causality and deployment-faithful evaluation pipelines. However, existing SSM approaches still rely on dense linear projections at the input and output layers,
which dominate parameter count and memory traffic.
Moreover, cross-KPI interactions are typically handled either implicitly or via lightweight heuristics,
leaving a gap between expressive multivariate modeling and strict Near-RT efficiency. Table~\ref{tab:related_work_overview} summarizes representative Transformer- and SSM-based models.

\begin{table}[t!]
\centering
\scriptsize
\setlength{\tabcolsep}{5pt}
\caption{Major Notations and Parameters.}
\label{tab:notation_qtn_s6}
\resizebox{\columnwidth}{!}{%
\begin{tabular}{lll}
\toprule
\textbf{Symbol} & \textbf{Description} & \textbf{Size/Type} \\
\midrule
\multicolumn{3}{l}{\emph{\textbf{Sets, operators, and basic math}}}\\
$\R,\ \mathbb{C}$ & Real and complex numbers & sets \\
$\E[\cdot],\ \Var[\cdot],\ \Cov[\cdot]$ & Expectation, variance, covariance & scalar/matrix \\
$\Tr(\cdot)$ & Trace & scalar \\
$\diag(\cdot)$ & Diagonal matrix from a vector & matrix \\
$\norm{\cdot},\ \abs{\cdot},\ \ip{\cdot}{\cdot}$ & Norm, absolute value, inner product & --- \\
$\odot$ & Hadamard (elementwise) product & --- \\
$\bigO(\cdot)$ & Big-\!O complexity notation & --- \\
$\PP,\ \cF$ & Probability measure; $\sigma$-algebra & --- \\
\midrule
\multicolumn{3}{l}{\emph{\textbf{Time, indices, and dimensions}}}\\
$t$ & Discrete time index & integer \\
$\Delta t$ & E2SM-KPM reporting interval / sampling period & scalar \\
$K$ & Number of monitored KPIs (features) & $\mathbb{Z}_+$ \\
$L$ & Lookback window length & $\mathbb{Z}_+$ \\
$H$ & Forecast horizon (in this paper $H{=}1$) & $\mathbb{Z}_+$ \\
$D$ & Latent width / number of channels & $\mathbb{Z}_+$ \\
$B_\ell$ & Number of stacked structured SSM blocks & $\mathbb{Z}_+$ \\
$N_s$ & Per-channel SSM state dimension & $\mathbb{Z}_+$ \\
$C_m$ & Number of HiPPO--LegS mixture components per block & $\mathbb{Z}_+$ \\
$D_m$ & ChannelMix hidden width (often $D_m=\alpha D$) & $\mathbb{Z}_+$ \\
$\alpha$ & ChannelMix expansion factor ($D_m=\alpha D$) & scalar \\
$N$ & Number of windows (samples) & $\mathbb{Z}_+$ \\
$B$ & Batch size & $\mathbb{Z}_+$ \\
$i,\ \ell,\ \tau$ & Feature index; time-in-window; lag index & integers \\
$t^\star$ & Target KPI index (e.g., RSRP) & index \\
\midrule
\multicolumn{3}{l}{\emph{\textbf{Telemetry, windows, and normalization}}}\\
$\mathbf{x}_t$ & KPI vector at time $t$ & $\R^{K}$ \\
$\mathbf{X}_t$ & Sliding input window ending at $t$ & $\R^{L\times K}$ \\
$\mathbf{u}_\ell$ & $\ell$-th row of $\mathbf{X}_t$ (one time step) & $\R^{K}$ \\
$y_t$ & Target KPI scalar at time $t$ (selected KPI) & $\R$ \\
$\hat{y}_{t+1}$ & Predicted next-step target & $\R$ \\
$\{(\mathbf{X}_n,\mathbf{y}_n)\}_{n=1}^N$ & Windowed dataset & $\mathbf{X}_n\!\in\!\R^{L\times K}$ \\
$\rho_{\mathrm{tr}},\rho_{\mathrm{va}}$ & Chronological split ratios (train/val) & scalars \\
$\mathcal{I}_{\mathrm{tr}},\mathcal{I}_{\mathrm{va}},\mathcal{I}_{\mathrm{te}}$ & Train/val/test index sets & subsets of $\{1{:}N\}$ \\
$\bm\mu_x,\bm\sigma_x$ & Train-only per-feature input mean/std & $\R^{K}$ \\
$\mu_y,\sigma_y$ & Train-only target mean/std & scalars \\
$\widetilde{\mathbf{X}},\widetilde{y}$ & Standardized inputs/target & as $\mathbf{X},y$ \\
$\epsilon$ & Small constant for std clamping & scalar \\
\midrule
\multicolumn{3}{l}{\emph{\textbf{Latent sequence, blocks, gating, and mixing}}}\\
$\mathbf{e}_\ell$ & Latent embedding at step $\ell$ & $\R^{D}$ \\
$\mathbf{E}$ & Latent sequence tensor (batch, time, channel) & $\R^{|\mathcal{B}|\times L\times D}$ \\
$\mathbf{Y},\mathbf{Z}$ & Intermediate block outputs (after conv/gate; after mix) & $\R^{|\mathcal{B}|\times L\times D}$ \\
$\mathbf{h}$ & Causal summary (last-step latent) & $\R^{|\mathcal{B}|\times D}$ \\
$\mathrm{LN}(\cdot)$ & Layer normalization & map $\R^{D}\!\to\!\R^{D}$ \\
$\mathrm{Drop}(\cdot)$ & Dropout operator & map \\
$p_{\mathrm{do}}$ & Dropout probability & scalar \\
$\sigma(\cdot)$ & Sigmoid nonlinearity (gating) & scalar-wise map \\
$\mathrm{GELU}(\cdot)$ & GELU nonlinearity (ChannelMix) & scalar-wise map \\
$\mathbf{g}$ & Squeeze-and-Excitation (SE) gate (channel gains) & $\R^{D}$ \\
$\mathbf{q}_\ell$ & ChannelMix gate at time step $\ell$ & $\R^{D_m}$ \\
$W_{\uparrow},W_{\downarrow}$ & ChannelMix up/down projections & $\R^{D\times 2D_m},\ \R^{D_m\times D}$ \\
\midrule
\multicolumn{3}{l}{\emph{\textbf{TT/MPS (Tensor Train) linear operators}}}\\
$\mathrm{TTIn}$ & TT-parameterized input projection $ \R^{K}\!\to\!\R^{D}$ & TTLinear \\
$\mathrm{TTHead}$ & TT-parameterized head $ \R^{D}\!\to\!\R$ & TTLinear \\
$\mathbf{n}^{\mathrm{in}},\mathbf{m}^{\mathrm{in}}$ & TT input in/out mode factorization & integer vectors \\
$r_{\mathrm{in}}$ & TT rank for input projection & $\mathbb{Z}_+$ \\
$\mathbf{n}^{\mathrm{hd}},\mathbf{m}^{\mathrm{hd}}$ & TT head in/out mode factorization & integer vectors \\
$r_{\mathrm{hd}}$ & TT rank for prediction head & $\mathbb{Z}_+$ \\
\midrule
\multicolumn{3}{l}{\emph{\textbf{HiPPO--LegS state-space kernels (per block/component)}}}\\
$A_{\mathrm{ct}}$ & Continuous-time HiPPO--LegS generator & $\R^{N_s\times N_s}$ \\
$B_{\mathrm{ref}}$ & Reference input vector for HiPPO construction & $\R^{N_s}$ \\
$\Delta t_0$ & Base initialization for time-step & scalar \\
$\gamma$ & Geometric growth factor across mixture components & scalar \\
$\Delta t^{(b,m)}$ & Learned positive step (block $b$, component $m$) & $\R_{>0}$ \\
$\bar A$ & Discretized transition (Tustin) & $\R^{N_s\times N_s}$ \\
$\bar B$ & Discretized input map (Tustin) & $\R^{D\times N_s}$ \\
$B^{(b,m)},C^{(b,m)}$ & Learned per-channel SSM parameters & $\R^{D\times N_s}$ \\
$D^{(b,m)}$ & Learned skip/diagonal term & $\R^{D}$ \\
$K^{(b,m)}$ & Component kernel taps (depthwise conv filter) & $\R^{D\times 1\times L}$ \\
$K^{(b)}$ & Mixture-summed kernel $K^{(b)}=\sum_m K^{(b,m)}$ & $\R^{D\times 1\times L}$ \\
$k[\tau]$ & Kernel tap at lag $\tau$ (per channel) & $\R^{D}$ \\
$\mathrm{conv1d}(\cdot)$ & Depthwise causal 1D convolution (groups=$D$) & operator \\
\midrule
\multicolumn{3}{l}{\emph{\textbf{Optimization, training loop, and inference metrics}}}\\
$\theta$ & Model parameters & collection \\
$\theta^\star$ & Best checkpoint parameters & collection \\
$E_{\max}$ & Maximum epochs & $\mathbb{Z}_+$ \\
$p$ & Early-stopping patience & $\mathbb{Z}_+$ \\
$\eta$ & Learning rate (AdamW) & scalar \\
$\lambda$ & Weight decay (AdamW) & scalar \\
$c_{\max}$ & Gradient clipping threshold & scalar \\
$\mathcal{L}_{\mathrm{tr}},\mathcal{L}_{\mathrm{va}}$ & Train/val loss (MSE in standardized space) & scalars \\
$\mathcal{L}^\star_{\mathrm{va}}$ & Best validation loss & scalar \\
\midrule
\multicolumn{3}{l}{\emph{\textbf{Complexity-relevant quantities}}}\\
$H_a$ & \# attention heads (Transformer baseline) & $\mathbb{Z}_+$ \\
$d_h$ & Per-head dimension ($D=H_a d_h$) & $\mathbb{Z}_+$ \\
\bottomrule
\end{tabular}%
}
\end{table}

\subsection{Contributions}
\label{sec:contributions}

Our contributions address the following technical gaps:

\begin{itemize}
  \item \textbf{Gap: quadratic (or near-quadratic) sequence cost in attention-based forecasting.}
  Transformer forecasters (e.g., Informer/FEDformer/ETSformer/PatchTST/Crossformer and recent variants) rely on attention or attention-like mechanisms, which introduce $\mathcal{O}(L^2)$ memory/compute terms or heavy parameterization that becomes costly as the lookback window $L$ grows and when deployed at scale in Near-RT xApps (Table~\ref{tab:related_work_overview}).
  \textbf{Solution:} We propose LiQSS, a post-Transformer design that removes self-attention entirely and instead uses structured state-space dynamics with linear-time and linear-memory dependence on $L$ (Section~\ref{subsec:complexity}).

  \item \textbf{Gap: weak dynamical inductive bias for radio telemetry under strict causality.}
  Many time-series Transformers treat telemetry as generic sequences and learn temporal structure primarily through attention, which is not explicitly aligned with stable continuous-time dynamics and can be insufficiently robust under distribution shifts and non-stationarity typical of radio KPIs (Section~\ref{sec:related_work}).
  \textbf{Solution:} We instantiate stable HiPPO--LegS state-space dynamics as causal depthwise convolutions, and enhance expressivity via a mixture of $C_m$ time scales per block (Algorithms~\ref{alg:s6_qtn_train_part1}--\ref{alg:s6_qtn_train_part2}, Section~\ref{subsec:ssm} and \ref{subsec:mixture}).

  \item \textbf{Gap: dense input/output projections dominate parameters and memory traffic in efficient sequence models.}
  Even when temporal backbones are linear-time (structured SSMs), global dense projections at the embedding and head often remain the main source of parameters and data movement, which is undesirable in Near-RT deployments (see SSM baselines in Table~\ref{tab:related_work_overview}).
  \textbf{Solution:} We compress the largest global operators by replacing dense embedding and readout maps with TT/MPS (Matrix Product Operator (MPO)~\cite{Guo2018MPOSeq2Seq, Ren2022TDDMRG_ComplexSystems}) factorizations (TTIn/TTHead), yielding rank-controlled global coupling with substantially fewer parameters (Section~\ref{subsec:tt}).

  \item \textbf{Gap: modeling cross-KPI interactions typically relies on expensive global attention.}
  Multivariate forecasting methods often use feature/temporal attention to capture cross-dimension dependence (e.g., Crossformer-style mechanisms), which increases runtime and footprint, and can be misaligned with tight latency budgets (Table~\ref{tab:related_work_overview}).
  \textbf{Solution:} We introduce lightweight interaction operators—squeeze--excitation gating (window-level reweighting) and token-wise gated ChannelMix (per-time-step coupling)—to capture non-stationary cross-KPI effects without temporal attention (Section~\ref{subsec:interaction}). 
\end{itemize}

The remainder of this paper is organized as follows. Section~\ref{subsec:qmb_primer} introduces the quantum many-body perspective that motivates the proposed approach.
Section~\ref{sec:system_model} describes the O-RAN system model, telemetry acquisition pipeline,
and Near-RT prediction task.
Section~\ref{sec:method} presents the proposed LiQSS
architecture, including its theoretical foundations, algorithmic formulation, and computational
complexity.
Section~\ref{sec:experimental-setup} details the experimental setup, datasets, baselines, and
evaluation protocol.
Section~\ref{sec:perf-numerical} reports and analyzes numerical results, accuracy--efficiency trade-offs,
and qualitative prediction behavior.
Finally, Section~\ref{sec:conclusion} concludes the paper. 
The major notations used throughout the paper are summarized in Table~\ref{tab:notation_qtn_s6}.

\section{Quantum Many-Body Primer}
\label{subsec:qmb_primer}

\subsection{From a Single Quantum System to Many-body Structure}
A quantum state of a single finite-dimensional system (a site) is a unit vector
$\ket{\psi}\in\mathbb{C}^{d}$. For a collection of $L$ sites, the joint (pure) state lives in the tensor-product
Hilbert space~\cite{BalazsTeofanov2022, Bach2023InformationTheoryKernel}

\begin{equation}
\begin{aligned}
\mathcal{H} = \bigotimes_{\ell=1}^{L}\mathbb{C}^{d_\ell},~
\ket{\psi} =
\sum_{i_1=1}^{d_1}\cdots\sum_{i_L=1}^{d_L}
\psi_{i_1,\dots,i_L}\,
\ket{i_1}\otimes\cdots\otimes\ket{i_L},
\end{aligned}
\label{eq:qmb_state}
\end{equation}
where the coefficient tensor $\Psi=[\psi_{i_1,\dots,i_L}]$ has $\prod_{\ell=1}^{L}d_\ell$ complex entries.
The exponential growth of degrees of freedom with $L$ is the central computational challenge of \emph{quantum many-body} systems~\cite{GlisicLorenzo2024QuantumNeuro6G7GSurvey, Tasaki2020QuantumManyBody}.

\subsection{Low-rank Structure via Entanglement}
Correlations in quantum mechanics are quantified by \emph{entanglement}~\cite{Cacciapuoti2024QuantumMAC}. Given a bipartition of sites into
$A=\{1,\dots,k\}$ and $B=\{k+1,\dots,L\}$, reshape $\Psi$ into a matrix
$\Psi_{(i_A),(i_B)}\in\mathbb{C}^{(\prod_{\ell\in A}d_\ell)\times(\prod_{\ell\in B}d_\ell)}$ and apply a Singular Value Decomposition (SVD)~\cite{ChengLiuEdforsVidal2025ParallelSVD}
\begin{equation}
\Psi_{(i_A),(i_B)}=\sum_{r=1}^{\chi_k} \sigma^{(k)}_r\, u^{(k)}_r(i_A)\, v^{(k)}_r(i_B)^{\!*},
\label{eq:schmidt}
\end{equation}
which is the \emph{Schmidt decomposition}~\cite{Kumar2024SchmidtProperties}. The Schmidt rank $\chi_k=\mathrm{rank}(\Psi_{(i_A),(i_B)})$ and singular values
$\{\sigma^{(k)}_r\}$ determine the bipartite entanglement entropy~\cite{CalabreseCardy2004EntanglementEntropyQFT}
\begin{equation}
S_k = -\sum_{r=1}^{\chi_k} p^{(k)}_r \log p^{(k)}_r,\qquad
p^{(k)}_r = \left(\sigma^{(k)}_r\right)^2 \Big/ \sum_{j=1}^{\chi_k}\left(\sigma^{(k)}_j\right)^2.
\label{eq:ent_entropy}
\end{equation}

Many physically relevant states exhibit \emph{limited} entanglement (e.g., area-law behavior~\cite{Paul2024AreaLawQuantumGeometry}), implying that
$\chi_k$ remains moderate for all cuts. This motivates representing $\Psi$ using factorizations whose intermediate
ranks correspond to $\chi_k$.

\subsection{Matrix Product States and Bond Dimension}
\label{subsec:operator}
A canonical low-entanglement representation is the MPS~\cite{Bradley2020MLST}, also known (in real-valued ML settings~\cite{Dborin2022MPSPretraining})
as the TT decomposition~\cite{Qiu2022EfficientTensorRPCA}:
\begin{equation}
\psi_{i_1,\dots,i_L}
=
\sum_{\alpha_1,\dots,\alpha_{L-1}}
G^{(1)}_{i_1,\alpha_1}
G^{(2)}_{\alpha_1,i_2,\alpha_2}
\cdots
G^{(L)}_{\alpha_{L-1},i_L},
\label{eq:mps}
\end{equation}
where $\alpha_\ell\in\{1,\dots,r_\ell\}$ are \emph{bond} indices and $\{r_\ell\}$ are \emph{bond dimensions}
(TT ranks). The bond dimensions upper-bound the Schmidt ranks across all bipartitions,
$\chi_k \le r_k$, and hence they control correlation capacity; larger ranks admit richer long-range dependencies, while
smaller ranks enforce a structured low-correlation prior.

\subsubsection{From States to Operators---MPO and TT-linear Maps}
\label{sub:tt_eq}
Forecasting models use linear maps (embeddings and readout) rather than quantum states. The operator analogue of MPS is
the MPO~\cite{Guo2018MPOSeq2Seq, Ren2022TDDMRG_ComplexSystems}, identical to a TT-parameterized matrix.
Given $W\in\mathbb{R}^{M\times N}$, reshape it into
$\mathcal{W}\in\mathbb{R}^{m_1\times\cdots\times m_d \times n_1\times\cdots\times n_d}$
with $M=\prod_{q=1}^{d}m_q$, $N=\prod_{q=1}^{d}n_q$, and write
\begin{equation}
\mathcal{W}(\mathbf{i},\mathbf{j})
=
\sum_{r_1,\dots,r_{d-1}}
\prod_{q=1}^{d} \mathcal{G}^{(q)}(r_{q-1}, i_q, j_q, r_q),~ r_0=r_d=1,
\label{eq:mpo}
\end{equation}
which matches the TT/MPS form used by \texttt{TTLinear} in our implementation. The TT rank thus plays the same
role as bond dimension in MPO-based quantum simulation; it limits the operator's ability to couple distant factor modes,
while drastically reducing parameters from $\mathcal{O}(MN)$ to $\mathcal{O}\!\left(\sum_q r_{q-1}m_qn_qr_q\right)$~\cite{DBLP:conf/nips/NovikovPOV15}.

\subsection{Where the ``Many-body'' Inspiration Enters our Algorithm}
Our model is \emph{quantum-inspired} in the following precise sense:
\begin{enumerate}[leftmargin=1.3em]
\item \textbf{Degrees of freedom (sites).} The latent representation at each time step,
$\mathbf{e}_\ell\in\mathbb{R}^{D}$, is interpreted as $D$ interacting degrees of freedom (analogous to sites in a
1D many-body chain). Temporal evolution is applied per degree of freedom via channel-wise state-space dynamics,
yielding a separable \emph{local} evolution component.
\item \textbf{Controlled interaction capacity.} Cross-channel interactions are introduced by lightweight coupling
operators (gating and gated mixing). These are intentionally low-order interaction mechanisms compared to dense
global coupling (full attention), mirroring the many-body principle of combining local dynamics with restricted
interaction structure.
\item \textbf{Tensor-network parameterization (MPS/MPO).} The \emph{largest} linear maps in the pipeline---the input
projection $\mathbb{R}^{K}\!\to\!\mathbb{R}^{D}$ and the readout $\mathbb{R}^{D}\!\to\!\mathbb{R}$---are constrained to
TT/MPS form (MPO factorization). This imports the key computational idea from many-body simulation; represent large
objects through a chain of small cores whose bond dimensions (TT ranks) provide an explicit, interpretable knob
for the accuracy--efficiency trade-off.
\end{enumerate}

In summary, the algorithm borrows the \emph{representational mechanism} of quantum many-body methods---low-rank
tensor networks with rank-controlled correlation capacity---and combines it with stable linear state-space dynamics
for temporal modeling. This yields a structured predictor whose expressivity is tunable (via TT ranks and state
dimension) while maintaining deployment-friendly complexity.

\begin{figure}[t]
\centering
\includegraphics[width=\linewidth]{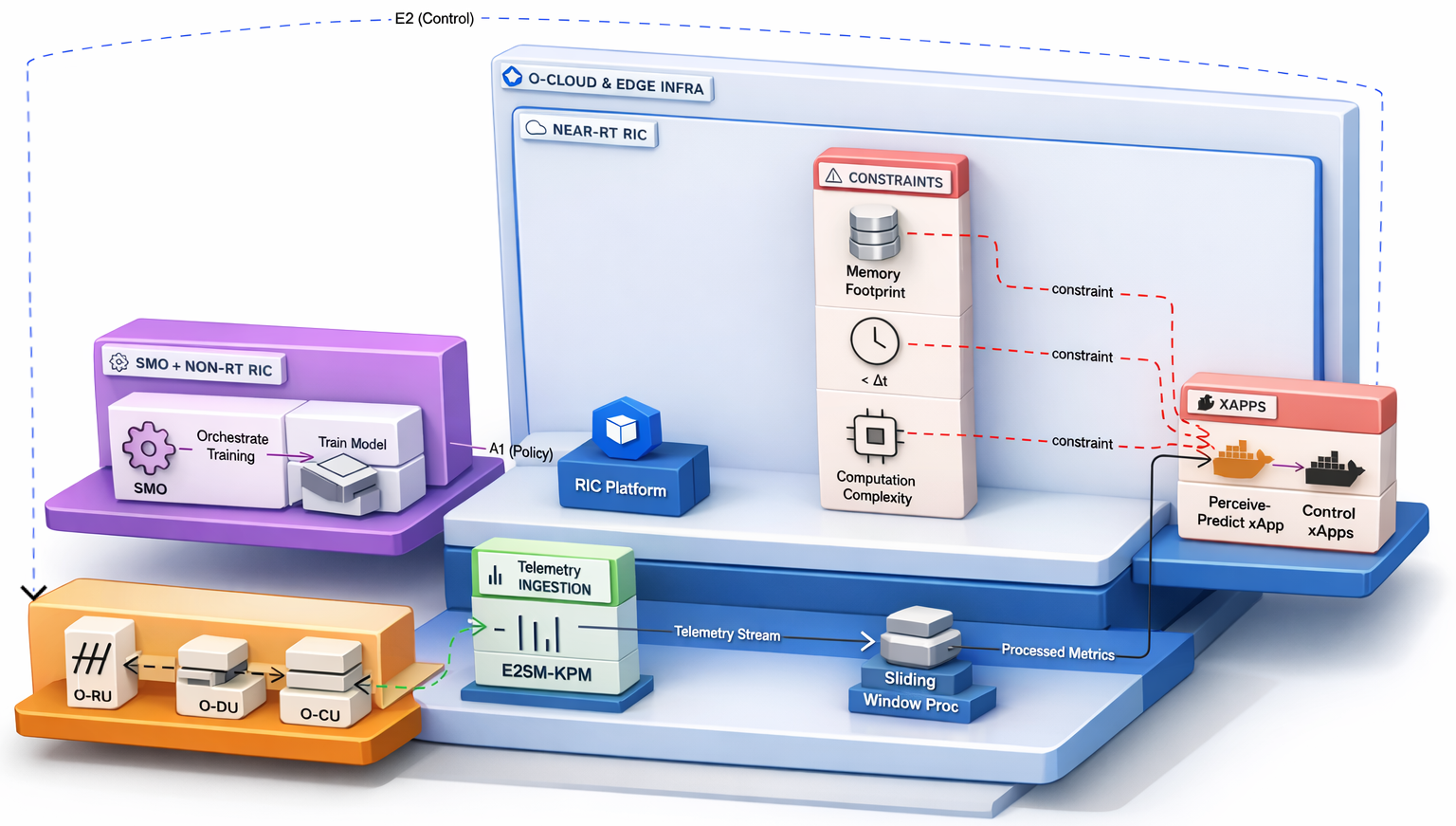}
\caption{System model and intelligence placement in the O-RAN architecture. Telemetry collected from E2 nodes (O-CU/O-DU) via the E2 interface using the E2SM-KPM service model is ingested by the Near-RT RIC and processed through sliding-window preprocessing. A perceive–predict xApp operates under Near-RT constraints on memory footprint, computational complexity, and latency ($<\Delta t$), producing short-horizon KPI predictions that drive near-real-time control actions via the E2 control loop. Long-term model training and policy generation are orchestrated by the SMO and Non-RT RIC, with resulting policies/model guidance delivered to the Near-RT RIC over the A1 interface, while trained models are deployed on O-Cloud/edge infrastructure for real-time operation.
}

\label{fig:system_workflow}
\end{figure}

\section{System Model}
\label{sec:system_model}

\subsection{O-RAN Functional Architecture and Intelligence Placement}
We consider the O-RAN Alliance architecture, which decomposes the RAN into interoperable and virtualized components with standardized interfaces~\cite{oran_architecture_overview, etsi_ts_104040}. Intelligence is introduced through the RICs, which are logically separated into a Non-Real-Time RIC (Non-RT RIC) and a Near-RT RIC, operating at different control time scales. Figure~\ref{fig:system_workflow} illustrates the considered O-RAN system model. The Non-RT RIC, hosted within the Service Management and Orchestration (SMO) framework, operates at time scales above one second and is responsible for long-term policy optimization, AI model training, lifecycle management, and enrichment of RAN data. The Near-RT RIC operates at time scales ranging from approximately $10$~ms to $1$~s and hosts latency-sensitive xApps that perform near-real-time analytics, prediction, and control. These xApps communicate with distributed RAN nodes (e.g., Distributed Unit (O-DU) and Central Unit (O-CU)) via the standardized E2 interface. In this work, we focus on intelligence deployed in the Near-RT RIC, where strict constraints on inference latency, computational complexity, and memory footprint limit the applicability of large-scale deep learning models.

\subsection{O-RAN Telemetry Acquisition via the E2 Interface}
The E2 interface provides standardized mechanisms for near-real-time telemetry reporting and control signaling between the Near-RT RIC and RAN nodes. Telemetry is exposed through E2 Service Models (E2SMs), such as E2SM-KPM, which define a structured framework for reporting Key Performance Measurements (KPMs) at configurable periodicities. Let $\Delta t$ denote the reporting interval configured for an E2SM-KPM subscription. At each reporting instant $t$, the Near-RT RIC receives KPI measurements aggregated over $\Delta t$ from one or multiple RAN entities. We model the received telemetry as a discrete-time multivariate signal
\begin{equation}
\mathbf{x}_t = \left[x_t^{(1)}, x_t^{(2)}, \dots, x_t^{(K)} \right]^\top \in \mathbb{R}^{K},
\end{equation}
where $K$ denotes the number of monitored KPIs. The telemetry stream is noisy and non-stationary due to user mobility, scheduling decisions, interference dynamics, and time-varying traffic.

\subsection{Near-RT Prediction Task Definition}
The objective of the perceive--predict xApp is to forecast short-horizon KPI values in order to enable proactive control decisions in the Near-RT RIC. Streaming telemetry received over the E2 interface is first segmented into overlapping sliding windows of fixed lookback length, which serve as strictly causal model inputs. In this work, we focus on \emph{single-step} forecasting ($H=1$) of a selected target KPI (e.g., RSRP), where the predictor maps each input window to a one-step-ahead estimate. The exact mathematical formulation of the sliding-window construction, causality constraints, and prediction function is given in Section~\ref{subsec:notation}. The formulation naturally extends to multi-step horizons ($H>1$) by modifying the output head and training labels, provided that inference latency remains within Near-RT bounds.

\subsection{Latency and Resource Constraints in the Near-RT RIC}
The Near-RT RIC is typically deployed on edge cloud infrastructure that is shared with other RAN functions and xApps, which imposes strict operational constraints on any embedded prediction model~\cite{almeida2023ric}. In particular, inference must be completed within a small fraction of the control loop interval $\Delta t$ in order to enable timely decision-making. At the same time, computational cost must scale linearly with the observation window length $L$ and the feature dimensionality to ensure scalability under increasing traffic and monitoring demands. Memory usage is also tightly constrained, requiring both model parameters and intermediate activations to fit within limited budgets available at the edge. Finally, excessive data movement between memory and compute units is undesirable, as virtualization and containerization overheads can significantly amplify memory traffic costs. Collectively, these constraints motivate the use of linear-complexity sequence models with compact parameterizations, rather than quadratic-cost attention-based architectures.

\begin{figure}[t]
\centering
\includegraphics[width=\linewidth]{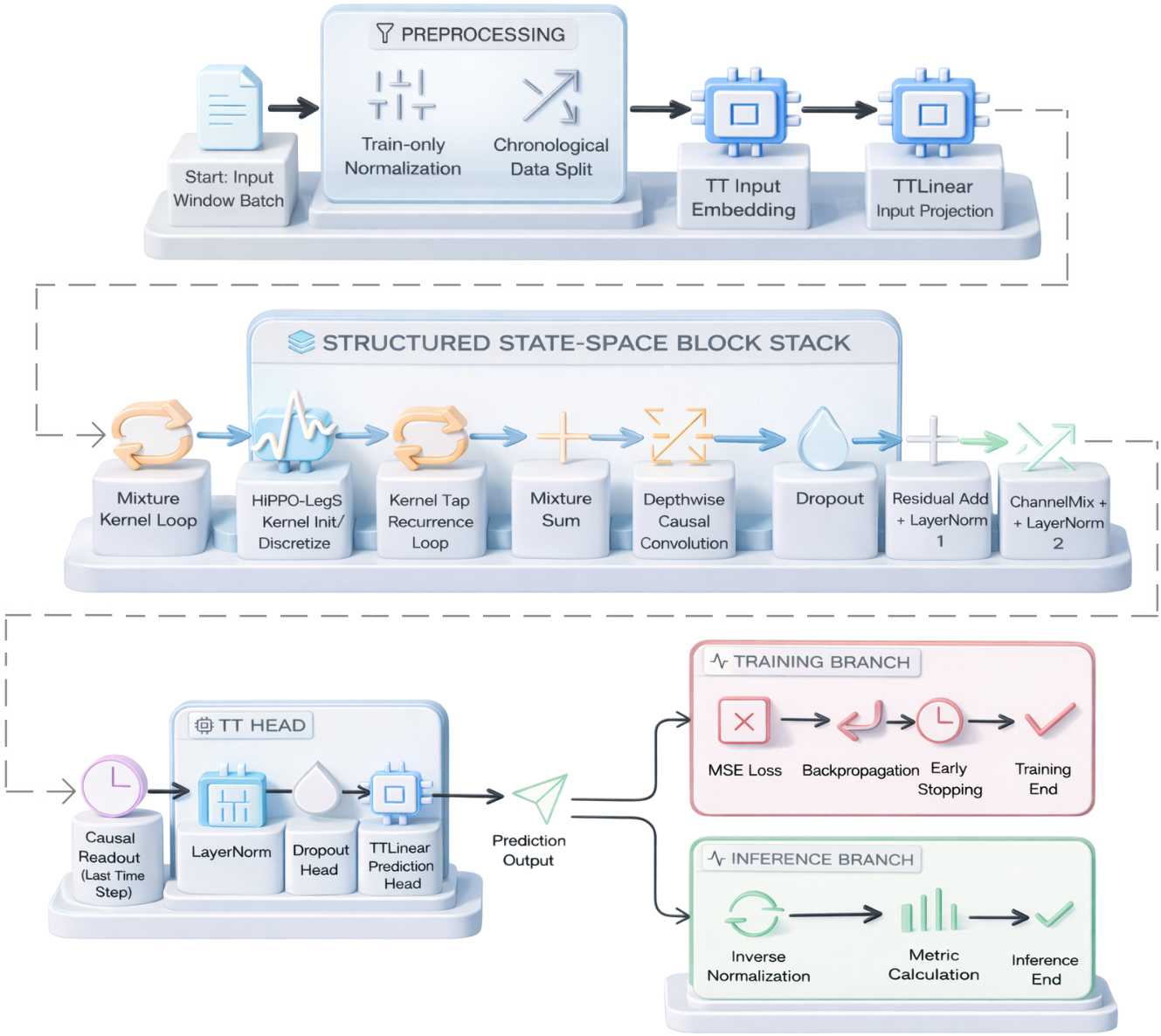}
\caption{General end-to-end architecture of the proposed quantum-inspired linear state-space tensor-network forecaster for Near-RT O-RAN telemetry. A chronologically split KPI window is standardized using \emph{train-set} statistics and embedded via a TT/MPS (TTLinear) input projection, then processed by a stack of structured HiPPO--LegS mixture state-space blocks implemented as depthwise causal convolutions with squeeze--excitation gating and gated channel mixing. A causal last-step readout summarizes the sequence, and a TT/MPS prediction head maps this causal summary to the next-step KPI estimate. The \emph{training} elements in the diagram denote the supervision/optimization loop applied to the model output (prediction $\hat{y}$ and ground truth $y$): parameters (including TT cores and SSM-mixture parameters) are learned by minimizing Mean Squared Error (MSE) $(\hat{y},y)$ via backpropagation with validation-driven early stopping and best-checkpoint saving. At \emph{inference}, the trained checkpoint is evaluated in a no-gradient forward pass, followed by inverse standardization (and metric computation when evaluating).}

\label{fig:system-modeling}
\end{figure}

\section{Quantum-Inspired Linear State-Space Tensor Network}
\label{sec:method}
Figure~\ref{fig:system-modeling} summarizes the end-to-end pipeline of the proposed \emph{LiQSS} predictor, from leakage-safe preprocessing to causal readout and deployment-faithful training/inference branches. In particular, the forward path replaces self-attention with structured HiPPO--LegS state-space dynamics realized as mixture-summed kernels and depthwise causal convolutions, while the largest global linear maps (input embedding and prediction head) are parameterized using TT/MPS operators. The complete, code-matched training procedure is given in Algorithm~\ref{alg:s6_qtn_train_part1} (setup, chronological splits, train-only normalization, and kernel construction) and Algorithm~\ref{alg:s6_qtn_train_part2} (end-to-end forward pass, optimization, validation, and early stopping). Algorithm~\ref{alg:s6_qtn_train_part2} is a direct continuation of Algorithm~\ref{alg:s6_qtn_train_part1}: it uses the leakage-safe scalers and the per-block \textsc{MixtureKernel} routine defined in Algorithm~\ref{alg:s6_qtn_train_part1} to execute the training loop and produce the best checkpoint parameters $\theta^\star$. The corresponding deployment-time inference, including inverse normalization and metric computation, is summarized in Algorithm~\ref{alg:s6_qtn_infer_codeexact}.

\subsection{Motivation and Conceptual Instantiation}
\label{subsec:foundation}

Based on the quantum many-body perspective introduced in the Section~\ref{subsec:qmb_primer}, we now clarify how these principles are \emph{instantiated} in the proposed architecture and why they are well aligned with Near-RT O-RAN constraints.

\subsubsection{From Many-body Structure to Model Design}
In quantum many-body systems, complex global behavior emerges from the combination of (i) local dynamics acting on individual degrees of freedom and (ii) restricted interaction operators whose expressive power is controlled by low-rank structure. Our design mirrors this principle at the algorithmic level.

After input embedding, each time step is represented by a latent vector
\begin{equation}
\mathbf{e}_\ell = [e^{(1)}_\ell,\dots,e^{(D)}_\ell]^\top \in \mathbb{R}^{D},
\end{equation}
whose coordinates are treated as distinct degrees of freedom. Temporal evolution is governed by channel-wise SSMs, which act independently on each latent dimension and therefore implement \emph{local} dynamics with linear-time complexity.

\subsubsection{Controlled Interaction Operators}
We introduce cross-channel dependencies through lightweight interaction mechanisms that intentionally avoid dense global coupling. Specifically:
\begin{itemize}
\item \emph{Squeeze--excitation gating} performs global but low-rank reweighting of channels based on pooled temporal statistics.
\item \emph{Gated channel mixing} introduces pointwise nonlinear coupling across channels without temporal entanglement.
\end{itemize}
These operators correspond to low-order interaction terms in a many-body system, capturing cross-KPI effects while preserving linear scaling in the window length.

\subsubsection{Tensor-network Parameterization of Global Maps}
The input embedding $\mathbb{R}^{K}\!\rightarrow\!\mathbb{R}^{D}$ and the prediction head $\mathbb{R}^{D}\!\rightarrow\!\mathbb{R}$ are implemented using TT/MPS-parameterized linear operators, ensuring that the most expensive global projections in the model respect the same structured, low-coupling design principles as the temporal dynamics.

\subsubsection{Block-level Composition}
Each processing block decomposes the sequence transformation into separable and non-separable components:
\begin{equation}
\mathbf{E}
\;\mapsto\;
\mathrm{LN}\!\left(
\mathbf{E}
+
\mathrm{Mix}\!\left(
\mathrm{Gate}\!\left(
\mathrm{SSM}(\mathbf{E})
\right)
\right)
\right),
\label{eq:foundation_block}
\end{equation}
where $\mathrm{SSM}(\cdot)$ encodes local temporal evolution, and $\mathrm{Gate}(\cdot)$ and $\mathrm{Mix}(\cdot)$ introduce controlled interactions. This separation is \emph{structural}, not heuristic, and ensures that long-range temporal modeling and cross-channel coupling are achieved without quadratic attention costs.


\subsection{Problem Setup and Notation}
\label{subsec:notation}

Let $\mathbf{x}_t \in \mathbb{R}^{K}$ denote the vector of $K$ radio KPIs observed at discrete time $t$. Given a sliding window of fixed length $L$, the input to the model is $\mathbf{X}_t = [\mathbf{x}_{t-L+1}, \dots, \mathbf{x}_t] \in \mathbb{R}^{L \times K}$. We consider \emph{single-step forecasting} ($H=1$) of a target KPI~\cite{Benidis2022DeepTimeSeriesSurvey, Salinas2020DeepAR}, $
\hat{y}_{t+1} = f_\theta(\mathbf{X}_t)$, where $f_\theta(\cdot)$ denotes the proposed predictor with parameters $\theta$.
For notational convenience, the $\ell$-th row of $\mathbf{X}_t$ is denoted by $\mathbf{u}_\ell \in \mathbb{R}^{K}$ for $\ell=1,\dots,L$. Chronological splitting is performed exactly as in Algorithm~\ref{alg:s6_qtn_train_part1}, lines~3--5.
Train-only normalization (inputs and target) is performed exactly as in Algorithm~\ref{alg:s6_qtn_train_part1}, lines~6--11, where statistics are computed on training slices only and then applied to all splits.


\subsection{Latent Representation Structure}
\label{subsec:manybody}

Each input vector $\mathbf{u}_\ell$ is mapped into a $D$-dimensional latent representation $\mathbf{e}_\ell$, as introduced in Section~\ref{subsec:foundation}. Latent channels are processed independently by channel-wise temporal dynamics and subsequently coupled through structured interaction operators (gating and mixing described in Section~\ref{subsec:interaction}). The latent tensor $\mathbf{E}\in\mathbb{R}^{|\mathcal{B}|\times L\times D}$ is produced by the TT input projection
(Algorithm~\ref{alg:s6_qtn_train_part2}, line~49) and subsequently updated in a block-wise manner by the structured state-space layers
(Algorithm~\ref{alg:s6_qtn_train_part2}, lines~50--63).



\begin{algorithm}[t]
\scriptsize
\caption{Setup \& Kernel}
\label{alg:s6_qtn_train_part1}
\begin{algorithmic}[1]
\Require Windows $\{(\mathbf{X}_n,\mathbf{y}_n)\}_{n=1}^N$ with $\mathbf{X}_n\!\in\!\mathbb{R}^{L\times K}$, $\mathbf{y}_n\!\in\!\mathbb{R}^{K}$;
target index $t^\star$ (RSRP); ratios $\rho_{\mathrm{tr}},\rho_{\mathrm{va}}$ (chronological slices);
batch size $B$; epochs $E_{\max}$; early-stop patience $p$;
LR $\eta$; weight decay $\lambda$; grad clip $c_{\max}{=}1.0$;
width $D$; layers $B_\ell$; state dim $N_s$; mixture comps $C_m$;
dropout $p_{\mathrm{do}}$; HiPPO base init $\Delta t_0$; dt growth $\gamma{=}1.5$;
TT in-modes $\mathbf{n}^{\mathrm{in}},\mathbf{m}^{\mathrm{in}}$, TT rank $r_{\mathrm{in}}$;
TT head modes $\mathbf{n}^{\mathrm{hd}},\mathbf{m}^{\mathrm{hd}}$, TT rank $r_{\mathrm{hd}}$;
device.
\Ensure Best params $\theta^\star$, scalers $(\bm\mu_x,\bm\sigma_x),(\mu_y,\sigma_y)$, best val loss $\mathcal{L}^\star_{\mathrm{va}}$.

\State \textbf{Harden gradients on.} \Comment{matches \texttt{torch.set\_grad\_enabled(True)}}
\State Enable global gradients.

\State \textbf{Chronological splits (slices).}
\State $n_{\mathrm{tr}}\gets\lfloor \rho_{\mathrm{tr}}N\rfloor$; $n_{\mathrm{va}}\gets\lfloor \rho_{\mathrm{va}}N\rfloor$.
\State $\mathcal{I}_{\mathrm{tr}}\gets\{1{:}n_{\mathrm{tr}}\}$;
$\mathcal{I}_{\mathrm{va}}\gets\{n_{\mathrm{tr}}{+}1{:}n_{\mathrm{tr}}{+}n_{\mathrm{va}}\}$;
$\mathcal{I}_{\mathrm{te}}\gets\{n_{\mathrm{tr}}{+}n_{\mathrm{va}}{+}1{:}N\}$.

\State \textbf{Train-only scalers (leakage-safe; exact).}
\State $\mathcal{X}_{\mathrm{tr}}\gets\{\mathbf{X}_n[\ell,:]\;|\;n\!\in\!\mathcal{I}_{\mathrm{tr}},\;\ell=1{:}L\}$.
\State $\bm\mu_x\gets\mathrm{mean}(\mathcal{X}_{\mathrm{tr}})$; $\bm\sigma_x\gets\mathrm{std}(\mathcal{X}_{\mathrm{tr}})$; clamp $\bm\sigma_x\ge\epsilon$.
\State $y_n\gets \mathbf{y}_n[t^\star]$ for all $n$.
\State $\mu_y\gets\mathrm{mean}(\{y_n\!:\!n\!\in\!\mathcal{I}_{\mathrm{tr}}\})$; $\sigma_y\gets\mathrm{std}(\{y_n\!:\!n\!\in\!\mathcal{I}_{\mathrm{tr}}\})$; clamp $\sigma_y\ge\epsilon$.
\State Scale all splits:
\[
\widetilde{\mathbf{X}}_n=(\mathbf{X}_n-\bm\mu_x)\oslash \bm\sigma_x,\qquad
\widetilde{y}_n=(y_n-\mu_y)/\sigma_y.
\]

\State \textbf{Model (code-exact modules).}
\State \textbf{TT input projection} $\mathrm{TTIn}:\mathbb{R}^{K}\to\mathbb{R}^{D}$ using TTLinear(in-modes $\mathbf{n}^{\mathrm{in}}$, out-modes $\mathbf{m}^{\mathrm{in}}$, rank $r_{\mathrm{in}}$, bias).
\State \textbf{structured state-space Blocks} $\{\mathcal{B}^{(b)}\}_{b=1}^{B_\ell}$.
Each block has mixture-kernel + depthwise causal conv + SE gate + dropout + LN + ChannelMix + LN.
\State \textbf{TT head} $\mathrm{TTHead}:\mathbb{R}^{D}\to\mathbb{R}$ using LayerNorm+Dropout+TTLinear(in-modes $\mathbf{n}^{\mathrm{hd}}$, out-modes $\mathbf{m}^{\mathrm{hd}}$, rank $r_{\mathrm{hd}}$).

\State \textbf{HiPPO--LegS fixed generator} (stored as buffer per component, numerically identical):
build $A_{\mathrm{ct}}\in\mathbb{R}^{N_s\times N_s}$, and reference $B_{\mathrm{ref}}\in\mathbb{R}^{N_s}$.

\Function{MixtureKernel}{$b$} \Comment{returns $K^{(b)}\in\mathbb{R}^{D\times 1\times L}$}
  \State $K^{(b)}\gets \mathbf{0}\in\mathbb{R}^{D\times 1\times L}$.
  \For{$m=1{:}C_m$} \Comment{each component is a HiPPOLegSKernel}
    \State \textbf{Component params (learned in code):} $B^{(b,m)}\!\in\!\mathbb{R}^{D\times N_s}$, $C^{(b,m)}\!\in\!\mathbb{R}^{D\times N_s}$, $D^{(b,m)}\!\in\!\mathbb{R}^{D}$, $\log\Delta t^{(b,m)}$.
    \State Init rule (code): $\Delta t^{(b,m)}_{\text{init}}=\Delta t_0\cdot \gamma^{(m-1)}$, stored via $\log\Delta t$.
    \State $\Delta t \gets \mathrm{softplus}(\log\Delta t^{(b,m)}) + 10^{-6}$.
    \State $I\gets \mathrm{Id}_{N_s}$; $\mathrm{lhs}\gets I-\tfrac{1}{2}\Delta t A_{\mathrm{ct}}$; $\mathrm{rhs}\gets I+\tfrac{1}{2}\Delta t A_{\mathrm{ct}}$.
    \State $\bar A \gets \mathrm{solve}(\mathrm{lhs},\mathrm{rhs})$.
    \State $\bar B \gets \mathrm{solve}(\mathrm{lhs},\Delta t\cdot B^{(b,m)\top})^\top$. \Comment{code computes via transpose}
    \State $\bar A_T \gets \bar A^\top$.
    \State \textbf{Recurrence taps (code exact).}
    \State $x \gets \bar B \in \mathbb{R}^{D\times N_s}$.
    \For{$\tau=0{:}L-1$}
      \State $k[\tau]\gets \sum_{j=1}^{N_s} C^{(b,m)}[:,j]\odot x[:,j] \in \mathbb{R}^{D}$.
      \State $x \gets x\,\bar A_T$.
    \EndFor
    \State $k[0]\gets k[0]+D^{(b,m)}$.
    \State Stack $K^{(b,m)}\in\mathbb{R}^{D\times L}$ from $\{k[\tau]\}_\tau$ and reshape to $K^{(b,m)}[:,1,:]\in\mathbb{R}^{D\times 1\times L}$.
    \State $K^{(b)}\gets K^{(b)} + K^{(b,m)}$. \Comment{mixture sum}
  \EndFor
  \State \Return $K^{(b)}$.
\EndFunction

\algstore{S6QTNTrainStore}
\end{algorithmic}
\end{algorithm}

\begin{algorithm}[t]
\scriptsize
\caption{Training}
\label{alg:s6_qtn_train_part2}
\begin{algorithmic}[1]
\algrestore{S6QTNTrainStore}

\State \textbf{Optimizer \& scheduler (code exact).}
\State AdamW with LR $\eta$ and weight decay $\lambda$.
\State ReduceLROnPlateau on val loss: factor $0.5$, patience $2$.
\State AMP autocast + GradScaler enabled iff CUDA.
\State Initialize $\mathcal{L}^\star_{\mathrm{va}}\gets +\infty$; remaining patience $\pi\gets p$; store checkpoint path.

\For{$e=1{:}E_{\max}$}
  \State \textbf{Train mode.} model $\gets$ \texttt{train}.
  \State \textbf{Training batches shuffled} (DataLoader \texttt{shuffle=True}).
  \State $\mathrm{sum}_{\mathrm{tr}}\gets 0$; $n_{\mathrm{tr}}\gets 0$.
  \State \textbf{Ensure grads on in loop} (\texttt{with torch.enable\_grad():}).

  \For{mini-batch $(\widetilde{\mathbf{X}}_{\mathcal{B}},\widetilde{\mathbf{y}}_{\mathcal{B}})$ from train loader}
    \State Move batch to device. Zero gradients (\texttt{set\_to\_none=True}).
    \State \textbf{Forward (vectorized; code exact).}
    \State $\mathbf{E}\gets \mathrm{TTIn}(\widetilde{\mathbf{X}}_{\mathcal{B}})\in\mathbb{R}^{|\mathcal{B}|\times L\times D}$.
    \For{$b=1{:}B_\ell$}
      \State $K^{(b)}\gets$ \Call{MixtureKernel}{$b$} \Comment{$D\times 1\times L$}
      \State \textbf{Depthwise causal conv (code exact):}
      \State $X_{dw}\gets \mathbf{E}^\top \in \mathbb{R}^{|\mathcal{B}|\times D\times L}$ (transpose time/channel).
      \State $Y_{dw}\gets \mathrm{conv1d}(\mathrm{pad\_left}(X_{dw},L-1),K^{(b)},\mathrm{groups}=D)$.
      \State $\mathbf{Y}\gets Y_{dw}^\top \in \mathbb{R}^{|\mathcal{B}|\times L\times D}$.
      \State \textbf{SE gate (code exact):} pool over time then MLP sigmoid $\Rightarrow$ channel gains; $\mathbf{Y}\gets \mathbf{Y}\odot \mathbf{g}$.
      \State \textbf{Dropout:} $\mathbf{Y}\gets \mathrm{Drop}(\mathbf{Y})$.
      \State \textbf{Residual + LN1:} $\mathbf{Y}\gets \mathrm{LN}_1(\mathbf{E}+\mathbf{Y})$.

      \State \textbf{ChannelMix (code exact: internal residual+LN).}
      \State $U\gets W_{\uparrow}\mathbf{Y}$; split $U=(\mathbf{a},\mathbf{g})$ along channels.
      \State $\mathrm{MLPout}\gets W_{\downarrow}(\mathrm{GELU}(\mathbf{a})\odot\sigma(\mathbf{g}))$.
      \State $\mathbf{Z}\gets \mathrm{LN}(\mathbf{Y}+\mathrm{Drop}(\mathrm{MLPout}))$. \Comment{ChannelMix returns this}

      \State \textbf{Extra residual + LN2 (code exact):} $\mathbf{E}\gets \mathrm{LN}_2(\mathbf{Y}+\mathbf{Z})$.
    \EndFor
    \State \textbf{Causal readout:} $\mathbf{h}\gets \mathbf{E}[:,L,:]$ (last time index).
    \State \textbf{TT head (code exact):} $\hat{\mathbf{y}}\gets \mathrm{TTHead}(\mathbf{h})\in\mathbb{R}^{|\mathcal{B}|\times 1}$.
    \State \textbf{Loss (scaled space):} $\mathcal{L}_{\mathcal{B}}\gets \mathrm{MSE}(\hat{\mathbf{y}},\widetilde{\mathbf{y}}_{\mathcal{B}})$.
    \State \textbf{Backward/update (AMP exact).}
    \State Scale loss; backprop; clip $\|\nabla\theta\|\le c_{\max}$; scaler step(AdamW); scaler update.
    \State $\mathrm{sum}_{\mathrm{tr}}\mathrel{+}= \mathcal{L}_{\mathcal{B}}\cdot|\mathcal{B}|$; $n_{\mathrm{tr}}\mathrel{+}=|\mathcal{B}|$.
  \EndFor
  \State $\mathcal{L}_{\mathrm{tr}}\gets \mathrm{sum}_{\mathrm{tr}}/\max(1,n_{\mathrm{tr}})$.

  \State \textbf{Validation mode.} model $\gets$ \texttt{eval}.
  \State $\mathrm{sum}_{\mathrm{va}}\gets 0$; $n_{\mathrm{va}}\gets 0$.
  \State Disable grads (\texttt{with torch.no\_grad():}).
  \For{mini-batch $(\widetilde{\mathbf{X}}_{\mathcal{B}},\widetilde{\mathbf{y}}_{\mathcal{B}})$ from val loader}
    \State $\hat{\mathbf{y}}\gets f_{\theta}(\widetilde{\mathbf{X}}_{\mathcal{B}})$ (same forward).
    \State $\mathcal{L}_{\mathcal{B}}\gets \mathrm{MSE}(\hat{\mathbf{y}},\widetilde{\mathbf{y}}_{\mathcal{B}})$.
    \State $\mathrm{sum}_{\mathrm{va}}\mathrel{+}= \mathcal{L}_{\mathcal{B}}\cdot|\mathcal{B}|$; $n_{\mathrm{va}}\mathrel{+}=|\mathcal{B}|$.
  \EndFor
  \State $\mathcal{L}_{\mathrm{va}}\gets \mathrm{sum}_{\mathrm{va}}/\max(1,n_{\mathrm{va}})$.
  \State Scheduler step on $\mathcal{L}_{\mathrm{va}}$.

  \If{$\mathcal{L}_{\mathrm{va}} < \mathcal{L}^\star_{\mathrm{va}} - 10^{-6}$}
    \State $\mathcal{L}^\star_{\mathrm{va}}\gets \mathcal{L}_{\mathrm{va}}$.
    \State Save checkpoint $\theta^\star\gets\theta$.
    \State $\pi\gets p$.
  \Else
    \State $\pi\gets \pi-1$.
    \If{$\pi=0$}
      \State \textbf{break}.
    \EndIf
  \EndIf
\EndFor

\State \textbf{Return} $\theta^\star$, $(\bm\mu_x,\bm\sigma_x)$, $(\mu_y,\sigma_y)$, $\mathcal{L}^\star_{\mathrm{va}}$.
\end{algorithmic}
\end{algorithm}

\subsection{Tensor-Network Linear Operators}
\label{subsec:tt}

The model employs Tensor Train (TT/MPS) parameterizations for selected linear maps, constraining global projections to structured operators while retaining code-level equivalence to dense linear layers at the API level. The TT input projection and TT head are specified in Algorithm~\ref{alg:s6_qtn_train_part1}, lines~13--15, and used in the training forward pass in Algorithm~\ref{alg:s6_qtn_train_part2}, lines~49 and~65, as well as in inference in Algorithm~\ref{alg:s6_qtn_infer_codeexact}, line~5.


\subsection{Input Embedding and Readout}
\label{subsec:embedding}


As introduced in Section~\ref{sub:tt_eq},
a TT/MPS (or MPO) representation factorizes a large linear operator into a chain of
small tensor cores whose bond dimensions control interaction capacity.

\subsubsection{TT-based Input Embedding}
\label{subsubsec:tt_in}

Let $\mathbf{u}_\ell\in\mathbb{R}^{K}$ denote the vector of KPIs observed at time step $\ell$.
The input embedding applies a TT-parameterized linear operator, $\mathrm{TTIn}:\mathbb{R}^{K}\rightarrow\mathbb{R}^{D}$, yielding the latent representation, $
\mathbf{e}_\ell = \mathrm{TTIn}(\mathbf{u}_\ell), \ell=1,\dots,L
\label{eq:tt_embedding}
$. In our implementation, we choose mode factorizations $\mathbf{n}^{\mathrm{in}}=(1,1,13)$ and
$\mathbf{m}^{\mathrm{in}}=(4,4,4)$ (so that $\prod \mathbf{n}^{\mathrm{in}}=13$ and $\prod \mathbf{m}^{\mathrm{in}}=64$),
and a TT rank $r_{\mathrm{in}}$, 
. This performs a structured global projection from KPI space into latent space, coupling inputs through a sequence
of low-rank interactions rather than a dense matrix.

\subsubsection{Causal Readout and TT-based Prediction Head}
\label{subsubsec:tt_head}

After processing the embedded sequence through the stacked structured state-space blocks,
a causal summary vector is extracted from the final time index, $\mathbf{h}_L = \mathbf{E}_{:\!,L,:} \in \mathbb{R}^{D}
\label{eq:causal_readout}$. This vector constitutes a sufficient statistic of the past window under the imposed causality constraint. The prediction is obtained by applying layer normalization followed by a TT-parameterized readout operator, $
\mathrm{TTHead}:\mathbb{R}^{D}\rightarrow\mathbb{R}
$, 
\begin{equation}
\hat{y}_{t+1}
=
\mathrm{TTHead}\!\left(\mathrm{LN}(\mathbf{h}_L)\right).
\label{eq:tt_readout}
\end{equation}

In our implementation, we use $\mathbf{n}^{\mathrm{hd}}=(4,4,4)$ and $\mathbf{m}^{\mathrm{hd}}=(1,1,1)$
(so that $\prod \mathbf{n}^{\mathrm{hd}}=64$ and $\prod \mathbf{m}^{\mathrm{hd}}=1$), and TT rank $r_{\mathrm{hd}}$. As in the input embedding, the TT factorization constrains the global readout map to a low-rank tensor-network structure,
ensuring that the final regression step remains parameter-efficient and consistent with the many-body inductive bias
of the model.



\subsection{Local Temporal Dynamics via HiPPO-LegS State-Space Models}
\label{subsec:ssm}

Each latent channel evolves according to an independent linear state-space model
\begin{align}
\mathbf{s}^{(d)}_{\ell+1} &= \mathbf{A}\mathbf{s}^{(d)}_{\ell} + \mathbf{b}^{(d)} e^{(d)}_\ell, \\
h^{(d)}_{\ell} &= (\mathbf{c}^{(d)})^\top \mathbf{s}^{(d)}_{\ell} + d^{(d)} e^{(d)}_\ell,
\end{align}
where $\mathbf{A}\in\mathbb{R}^{N_s\times N_s}$ is shared across channels, while $\mathbf{b}^{(d)}, \mathbf{c}^{(d)}, d^{(d)}$ are learned per-channel parameters.

\subsubsection{HiPPO-LegS Parameterization}
The continuous-time generator $\mathbf{A}_{\mathrm{ct}}$ is defined by the HiPPO-LegS~\cite{gu2022hippo} operator. A stable discrete-time transition matrix is obtained via bilinear (Tustin) discretization~\cite{Hespanha2018LinearSystems}
\begin{equation}
\mathbf{A} =
\left(\mathbf{I}-\frac{\Delta t}{2}\mathbf{A}_{\mathrm{ct}}\right)^{-1}
\left(\mathbf{I}+\frac{\Delta t}{2}\mathbf{A}_{\mathrm{ct}}\right).
\end{equation}

\subsubsection{Convolutional Form}
The resulting system admits a causal convolution representation
\begin{equation}
h^{(d)}_{\ell} = \sum_{\tau=0}^{\ell-1}
(\mathbf{c}^{(d)})^\top \mathbf{A}^{\tau}\mathbf{b}^{(d)} \,
e^{(d)}_{\ell-\tau}.
\end{equation}

The HiPPO-LegS fixed generator $(A_{\mathrm{ct}},B_{\mathrm{ref}})$ is constructed as in Algorithm~\ref{alg:s6_qtn_train_part1}, line~16.
The bilinear discretization is performed as in Algorithm~\ref{alg:s6_qtn_train_part1}, lines~22--25.
Kernel tap generation via the state recurrence is performed as in Algorithm~\ref{alg:s6_qtn_train_part1}, lines~27--33.
The resulting kernel is then applied by depthwise causal convolution in Algorithm~\ref{alg:s6_qtn_train_part2}, lines~52--55.


\subsection{Mixture of HiPPO-LegS Kernels}
\label{subsec:mixture}

To increase expressivity while preserving linear complexity, each block employs a mixture of $C_m$ state-space kernels, $\mathbf{K} = \sum_{m=1}^{C_m} \mathbf{K}^{(m)}$. Each component corresponds to an independent HiPPO-LegS parameterization with its own learned $(B,C,D)$ and time scale $\Delta t$. Mixture construction is defined by the loop over components (Algorithm~\ref{alg:s6_qtn_train_part1}, lines~19--35),
including component initialization (line~21), discretization (lines~22--25), recurrence (lines~27--33), and mixture summation (line~34).
Kernel creation is invoked per block during training (Algorithm~\ref{alg:s6_qtn_train_part2}, line~51).

\begin{algorithm}[t]
\scriptsize
\caption{Inference}
\label{alg:s6_qtn_infer_codeexact}
\begin{algorithmic}[1]
\Require Best params $\theta^\star$; scalers $(\bm\mu_x,\bm\sigma_x)$ and $(\mu_y,\sigma_y)$;
test loader providing scaled $(\widetilde{\mathbf{X}}_{\mathcal{B}},\widetilde{\mathbf{y}}_{\mathcal{B}})$.
\Ensure Predictions $\hat{y}$; MSE, RMSE, MAE in original units.

\State Load $\theta^\star$; set model to \texttt{eval}.
\State $\mathrm{SSE}\gets 0$, $\mathrm{SAE}\gets 0$, $M\gets 0$.
\State Disable grads (\texttt{with torch.no\_grad():}).

\For{mini-batch $(\widetilde{\mathbf{X}}_{\mathcal{B}},\widetilde{\mathbf{y}}_{\mathcal{B}})$ from test loader}
  \State \textbf{Forward (in scaled space).} $\widetilde{\hat{\mathbf{y}}}\gets f_{\theta^\star}(\widetilde{\mathbf{X}}_{\mathcal{B}})$.
  \State \textbf{Inverse-scale (code exact).}
  \State $\hat{\mathbf{y}}\gets \sigma_y\,\widetilde{\hat{\mathbf{y}}}+\mu_y$.
  \State $\mathbf{y}^{\mathrm{orig}}\gets \sigma_y\,\widetilde{\mathbf{y}}_{\mathcal{B}}+\mu_y$ \Comment{since loader stores scaled targets}
  \State $\mathrm{SSE}\mathrel{+}= \sum(\hat{\mathbf{y}}-\mathbf{y}^{\mathrm{orig}})^2$;\;
        $\mathrm{SAE}\mathrel{+}= \sum|\hat{\mathbf{y}}-\mathbf{y}^{\mathrm{orig}}|$.
  \State $M\mathrel{+}=$ number of scalar elements in $\hat{\mathbf{y}}$.
\EndFor

\State $\mathrm{MSE}\gets \mathrm{SSE}/\max(1,M)$;
$\mathrm{RMSE}\gets\sqrt{\mathrm{MSE}}$;
$\mathrm{MAE}\gets \mathrm{SAE}/\max(1,M)$.
\State \Return predictions and metrics.
\end{algorithmic}
\end{algorithm}

\subsection{Channel Gating and Mixing}
\label{subsec:interaction}

After the depthwise causal state-space convolution, each block produces an intermediate latent sequence
$\mathbf{Y}\in\mathbb{R}^{B\times L\times D}$, where $B$ is the batch size, $L$ the window length,
and $D$ the latent channel width. To capture non-stationary cross-channel (cross-KPI) dependencies without
self-attention, we apply two lightweight interaction operators:
(i) squeeze--excitation (SE) channel gating, and
(ii) token-wise gated channel mixing.
Both operators preserve linear scaling in $L$ and are composed using residual connections and layer normalization.

\subsubsection{Squeeze--excitation channel gating}
\label{subsubsec:segate_math}

SE gating introduces global, window-level channel reweighting based on aggregated temporal statistics.
First, a per-channel summary is obtained by averaging over the temporal dimension
\begin{equation}
\mathbf{s}
\;=\;
\frac{1}{L}\sum_{\ell=1}^{L}\mathbf{Y}_{:\!,\ell,:}
\in\mathbb{R}^{B\times D}.
\label{eq:se_pool}
\end{equation}

This summary is mapped to channel gains through a low-rank gating operator
\begin{equation}
\mathbf{g}
\;=\;
\sigma\!\left(\phi\!\left(\mathbf{s}W_1+\mathbf{b}_1\right)W_2+\mathbf{b}_2\right)
\in(0,1)^{B\times D},
\label{eq:se_gate}
\end{equation}
where $\phi(\cdot)$ is a pointwise nonlinearity, $\sigma(\cdot)$ is the sigmoid function,
$W_1\in\mathbb{R}^{D\times D_r}$, $W_2\in\mathbb{R}^{D_r\times D}$, and
$D_r=\max(1,\lfloor D/r\rfloor)$ is a reduction dimension.
The gated sequence is then obtained by channel-wise modulation
\begin{equation}
\widetilde{\mathbf{Y}}_{:\!,\ell,:}
\;=\;
\mathbf{Y}_{:\!,\ell,:}\odot \mathbf{g},
\qquad \ell=1,\dots,L,
\label{eq:se_apply}
\end{equation}
where $\odot$ denotes elementwise multiplication with broadcasting over time.
SE gating acts as a structured global interaction operator, adaptively emphasizing informative channels
based on the current temporal context while avoiding quadratic-time attention.

\subsubsection{Token-wise gated channel mixing}
\label{subsubsec:channelmix_math}

To introduce nonlinear cross-channel interactions at each time step without mixing information across time,
we apply a gated channel-mixing operator independently at each temporal index.
Let $\mathbf{u}_\ell\in\mathbb{R}^{B\times D}$ denote the input at time step $\ell$.
First, an expansion and split operation produces content and gate components
\begin{equation}
\big[\mathbf{a}_\ell,\mathbf{q}_\ell\big]
=
\mathbf{u}_\ell W_{\uparrow}+\mathbf{b}_{\uparrow},
\qquad
\mathbf{a}_\ell,\mathbf{q}_\ell\in\mathbb{R}^{B\times D_m},
\label{eq:mix_up}
\end{equation}
where $W_{\uparrow}\in\mathbb{R}^{D\times 2D_m}$ and $D_m=\alpha D$ is an expansion dimension.
The mixed output is then formed by gated nonlinear interaction followed by dimensionality reduction
\begin{equation}
\mathrm{Mix}(\mathbf{u}_\ell)
=
\Big(\phi(\mathbf{a}_\ell)\odot\sigma(\mathbf{q}_\ell)\Big)W_{\downarrow}+\mathbf{b}_{\downarrow},
\qquad
W_{\downarrow}\in\mathbb{R}^{D_m\times D}.
\label{eq:mix_down}
\end{equation}

This operator couples channels within each time step while preserving temporal independence across $\ell$.

\subsubsection{Residual composition at the block level}
\label{subsubsec:block_residuals}

Let $\mathbf{E}\in\mathbb{R}^{B\times L\times D}$ denote the input to a block, and
$\mathbf{Y}=\mathrm{SSM}(\mathbf{E})$ the output of the structured state-space operator.
The block output is formed through the following sequence of residual compositions
\begin{align}
\mathbf{Y}_1
&=
\mathrm{LN}_1\!\Big(\mathbf{E}+\mathrm{Drop}\big(\mathrm{SE}(\mathbf{Y})\big)\Big),
\label{eq:block_ln1}\\[4pt]
\mathbf{Z}
&=
\mathrm{LN}_m\!\Big(\mathbf{Y}_1+\mathrm{Drop}\big(\mathrm{Mix}(\mathbf{Y}_1)\big)\Big),
\label{eq:block_mix}\\[4pt]
\mathbf{E}^{+}
&=
\mathrm{LN}_2\!\Big(\mathbf{Y}_1+\mathbf{Z}\Big).
\label{eq:block_ln2}
\end{align}

The first residual path injects temporally filtered features into the block input.
The second applies gated channel mixing with normalization.
The final outer residual introduces an additional skip connection around the entire interaction sublayer,
improving optimization stability and enabling deep stacking of blocks.
Together, these operations provide controlled cross-channel interaction while maintaining linear-time
temporal modeling and strict causality.


\subsection{End-to-End Forward Map}
\label{subsec:forward_steps}

For a mini-batch in scaled space, the forward pass proceeds as follows:
\begin{enumerate}
\item \textbf{TT input embedding:} $\mathbf{E}\leftarrow \mathrm{TTIn}(\widetilde{\mathbf{X}}_{\mathcal{B}})$.
\item \textbf{Block stack:} for $b=1{:}B_\ell$, construct $K^{(b)}$ and apply depthwise causal convolution, gating, and channel mixing.
\item \textbf{Causal readout:} $\mathbf{h}\leftarrow \mathbf{E}[:,L,:]$ (last time index).
\item \textbf{TT head:} $\hat{\mathbf{y}}\leftarrow \mathrm{TTHead}(\mathbf{h})$.
\end{enumerate}

These steps correspond exactly to Algorithm~\ref{alg:s6_qtn_train_part2}, lines~49--65, with kernel construction at line~51 and depthwise causal convolution at lines~52--55.


\subsection{Training and Inference}
\label{subsec:train_infer}

The model is trained using mean-squared error in normalized space and optimized with AdamW, a validation-driven learning-rate scheduler, mixed precision when available, gradient clipping, and early stopping. Optimizer, scheduler, and Automatic Mixed Precision (AMP) configuration follow Algorithm~\ref{alg:s6_qtn_train_part2}, lines~36--40. The training loop (including device transfer, zeroing gradients, forward, loss, backward, clipping, and update) follows lines~41--69. Validation follows lines~71--79, and early stopping/checkpointing follows lines~80--87, returning the best checkpoint at line~88. Inference initialization follows Algorithm~\ref{alg:s6_qtn_infer_codeexact}, lines~1--3.
Forward prediction in scaled space is performed at line~5.
Inverse-scaling is performed at lines~7--8, error accumulation at line~9--10, metric computation at line~11, and return at line~12.

\subsection{Computational Complexity}
\label{subsec:complexity}

We analyze the computational and memory complexity of the proposed LiQSS architecture and compare it with standard Transformer self-attention. The analysis is expressed in terms of the window length $L$, latent width $D$, state dimension $N_s$, and number of stacked blocks $B_\ell$. Each structured state-space block performs channel-wise temporal modeling via causal depthwise convolutions, followed by lightweight interaction operators (squeeze--excitation gating and gated channel mixing). For a single block, the dominant cost arises from state-space kernel evaluation and token-wise channel mixing. Aggregating over $B_\ell$ blocks, the per-window computational complexity is $
\mathcal{O}\!\left(B_\ell\,L\,D\,N_s\right)
$, which scales linearly with the sequence length $L$ and contains no quadratic terms. Activation memory likewise scales linearly in $L$, with an additional $\mathcal{O}(D N_s)$ contribution from state buffers. By construction, the model avoids attention matrices entirely, eliminating the $\mathcal{O}(L^2)$~\cite{vaswani2017attention} compute and memory overhead associated with full self-attention. Furthermore, Tensor Train (TT/MPS) parameterization of the input projection and prediction head substantially reduces parameter count and memory traffic for global linear maps without altering asymptotic runtime complexity. To facilitate comparison, consider a sequence model with input length $L$ and latent width $D$. Suppose that:
(i) a Transformer layer employs full self-attention with $H_a$ heads and per-head dimension $d_h$ (so $D = H_a d_h$), and
(ii) the proposed model uses channel-wise state-space dynamics with state dimension $N_s$ and a gated channel-mixing layer of width $D_m = \alpha D$, $\alpha \ge 1$.
Ignoring constant factors and bias terms, the per-layer complexity is summarized below.
\begin{mdframed}[
  backgroundcolor=gray!5,
  linecolor=gray!50,
  linewidth=0.5pt,
  roundcorner=2pt,
  innerleftmargin=8pt,
  innerrightmargin=8pt,
  innertopmargin=6pt,
  innerbottommargin=6pt,
  skipabove=6pt,
  skipbelow=6pt,
  userdefinedwidth=\columnwidth,
  align=center
]
\footnotesize
\begin{enumerate}
\item \textbf{Transformer (full self-attention), per layer:}
\[
\begin{aligned}
\text{time} \;&=\; \mathcal{O}\!\left(L D^2 + L^2 D\right), \\
\text{activation memory} \;&=\; \mathcal{O}\!\left(L^2\right).
\end{aligned}
\]

\item \textbf{Proposed SSM + TT block, per layer:}
\[
\begin{aligned}
\text{time} \;&=\; \mathcal{O}\!\left(L D N_s + L D D_m\right), \\
\text{activation memory} \;&=\; \mathcal{O}\!\left(L D + D N_s\right).
\end{aligned}
\]
\end{enumerate}

\scriptsize
For sufficiently large $L$ and fixed $D$ and $N_s$, Transformer layers exhibit quadratic scaling in $L$, whereas the proposed architecture scales linearly. Here, \emph{time complexity} refers to asymptotic operation count and is hardware-agnostic, while \emph{activation memory} denotes runtime storage of intermediate tensors during forward and backward passes.
\end{mdframed}

\section{Experimental Setup}
\label{sec:experimental-setup}

\subsection{Data Collection and KPIs}
\label{subsec:data_collection_kpis}
The end-to-end telemetry acquisition pipeline and the associated implementation details follow the same
procedure established in our earlier studies~\cite{dai2025orankpi, Rezazadeh2025ReservoirAugmented}.
In brief, KPIs are collected from an O-RAN test environment and curated into a forecasting-ready dataset under a
consistent preprocessing and quality-control workflow.
To facilitate reproducibility and encourage further investigation, we release the dataset publicly\footnote{A curated release is hosted on IEEE DataPort: \url{https://ieee-dataport.org/documents/video-streaming-network-kpis-o-ran-testing}}. We consider $K=13$ KPIs extracted from the E2 Service Model for E2SM-KPM reports: Modulation and Coding Scheme (MCS), Channel Quality Indicator (CQI), Rank Indicator (RI), Precoding Matrix Indicator (PMI), Buffer occupancy (Buffer), Physical Resource Blocks (PRBs), Reference Signal Received Quality (RSRQ),  RSRP, Received Signal Strength Indicator (RSSI), SINR, Spectral Efficiency, Block Error Rate (BLER), and Delay.
The primary forecasting task in this paper is next-step prediction of \emph{RSRP} (target index $t^\star$), a representative radio-quality KPI widely used for mobility management and link-adaptation decisions.

\subsection{Baselines}
\label{subsec:baselines}

To ensure a fair and representative evaluation, we compare the proposed LiQSS model against a diverse set of
state-of-the-art baselines spanning Transformer-based architectures, attention-efficient variants, foundation-style time-series models, and structured state-space models. All selected baselines are widely used in the time-series literature or have been recently proposed
for long-sequence modeling, and collectively reflect the design space summarized in
Table~\ref{tab:related_work_overview} of Section~\ref{sec:related_work}. All baselines are evaluated under the same experimental protocol:
chronological data splits, train-only normalization, strictly causal inference,
and identical evaluation metrics.
Inference latency is measured on GPU using synchronized CUDA events~\cite{nvidia_cuda_events,pytorch_cuda_event},
ensuring fair comparison of deployment-relevant performance. For a fair comparison under identical time and memory constraints, we standardize the data pipeline and training procedure across all models. We also use compact baseline versions that retain each method’s key inductive bias. These implementations are not faithful re-runs of the original codebases; they are designed for consistency and comparability rather than exact replication.

\begin{table}[t!]
\centering
\footnotesize
\setlength{\tabcolsep}{13pt}
\caption{LiQSS Model Hyperparameter Configuration.}
\label{tab:hparams_liqss}
\resizebox{\columnwidth}{!}{%
\begin{tabular}{lll}
\toprule
\textbf{Symbol} & \textbf{Description} & \textbf{Value} \\
\midrule
\multicolumn{3}{l}{\emph{\textbf{Input, windows, and dimensions}}} \\
$K$ & Number of monitored KPIs & $13$ \\
$L$ & Lookback window length & 32 \\
$H$ & Forecast horizon & $1$ \\
$D$ & Latent channel width & $64$ \\
$B_\ell$ & Number of stacked SSM blocks & $2$ \\
$N_s$ & Per-channel SSM state dimension & $32$ \\
$C_m$ & HiPPO--LegS mixture components per block & $2$ \\
$\alpha$ & ChannelMix expansion factor & $1$ \\
$D_m$ & ChannelMix hidden width ($D_m=\alpha D$) & $64$ \\
\midrule
\multicolumn{3}{l}{\emph{\textbf{Regularization and gating}}} \\
$p_{\mathrm{do}}$ & Dropout probability & $0.1$ \\
\midrule
\multicolumn{3}{l}{\emph{\textbf{HiPPO--LegS discretization}}} \\
$\Delta t_0$ & Base time-step initialization & $0.1$ \\
$\gamma$ & Geometric growth factor & $1.5$ \\
$\Delta t^{(b,m)}$ & Learned positive step size & learned \\
\midrule
\multicolumn{3}{l}{\emph{\textbf{TT/MPS parameterization}}} \\
$\mathbf{n}^{\mathrm{in}}$ & TT input mode factorization & $(1,1,13)$ \\
$\mathbf{m}^{\mathrm{in}}$ & TT input output modes & $(4,4,4)$ \\
$r_{\mathrm{in}}$ & TT rank (input projection) & $4$ \\
$\mathbf{n}^{\mathrm{hd}}$ & TT head input modes & $(4,4,4)$ \\
$\mathbf{m}^{\mathrm{hd}}$ & TT head output modes & $(1,1,1)$ \\
$r_{\mathrm{hd}}$ & TT rank (prediction head) & $4$ \\
\midrule
\multicolumn{3}{l}{\emph{\textbf{Optimization and training}}} \\
$\eta$ & Learning rate (AdamW) & $3\times10^{-3}$ \\
$\lambda$ & Weight decay (AdamW) & $10^{-4}$ \\
$c_{\max}$ & Gradient clipping threshold & $1.0$ \\
$E_{\max}$ & Maximum training epochs & $120$ \\
$p$ & Early-stopping patience & $30$ \\
$B$ & Batch size & $256$ \\
$\rho_{\mathrm{tr}}$ & Training split ratio & $0.70$ \\
$\rho_{\mathrm{va}}$ & Validation split ratio & $0.15$ \\
\bottomrule
\end{tabular}%
}
\end{table}

\subsection{Model Hyperparameters}
\label{subsec:model_hparams}
All experiments use a fixed, code-faithful hyperparameter configuration summarized in Table~\ref{tab:hparams_liqss} (notations follow Table~\ref{tab:notation_qtn_s6}). Beyond these hyperparameters, experiments use chronological data splitting with the remaining $15\%$ reserved for testing, normalize features using training-set statistics only, include bias terms in both TT operators, learn $\Delta t^{(b,m)}$ with a softplus positivity constraint while keeping the continuous-time generator fixed, use a fixed random seed of $42$, and run on a single CUDA-enabled NVIDIA T4 GPU.

\begin{table*}[t!]
\centering
\scriptsize
\setlength{\tabcolsep}{13pt}
\caption{Benchmark Results Comparing the Proposed LiQSS with SSM- and Transformer-Based Baselines.}
\label{tab:results-liqss}
\resizebox{\textwidth}{!}{%
\begin{tabular}{lrrrrrrrr}
\toprule
\textbf{Model} & \textbf{RMSE} & \textbf{MAE} & \textbf{MSE} & \textbf{Skill (R)} & \textbf{Skill (M)} & $\boldsymbol{R^2}$ & \textbf{\#Params} & \textbf{Infer (s)} \\
\midrule
\multicolumn{9}{l}{\textbf{Proposed Method}} \\
\midrule
LiQSS &
0.286599 & 0.164083 & 0.082139 &
0.919865 & 0.939650 & 0.993276 &
44{,}109 & 0.000456 \\
\midrule
\multicolumn{9}{l}{\textbf{SSM-based Methods}} \\
\midrule
WM--MS$^{3}$M~\cite{rezazadeh2025agentic} &
0.291687 & 0.167417 & 0.085081 &
0.918448 & 0.938430 & 0.993035 &
477{,}802 & 0.000647 \\
MS$^{3}$M \cite{rez2025rivaling} &
0.291806 & 0.170298 & 0.085151 &
0.918414 & 0.937370 & 0.993030 &
698{,}449 & 0.000643 \\
\midrule
\multicolumn{9}{l}{\textbf{Transformer-based Methods}} \\
\midrule
RWKV \cite{peng2023rwkv} &
1.480259 & 0.407525 & 2.191166 &
0.576394 & 0.848496 & 0.812361 &
26{,}269{,}696 & 0.002103 \\
Performers \cite{choromanski2021rethinking} &
0.786704 & 0.276759 & 0.618902 &
0.774869 & 0.897110 & 0.947001 &
14{,}156{,}160 & 0.002634 \\
RetNet \cite{sun2023retentive} &
0.598374 & 0.208329 & 0.358051 &
0.828763 & 0.922550 & 0.969338 &
18{,}930{,}432 & 0.001522 \\
Chronos-T5 \cite{ansari2024chronos} &
2.518750 & 1.673298 & 6.344100 &
0.279208 & 0.377925 & 0.456727 &
14{,}209{,}536 & 0.000790 \\
Chronos-GPT \cite{ansari2024chronos} &
0.445510 & 0.183071 & 0.198479 &
0.872508 & 0.931941 & 0.983003 &
6{,}840{,}320 & 0.000749 \\
PatchTST \cite{nie2023patchtst} &
3.257556 & 2.539881 & 10.611672 &
0.089299 & 0.066044 & 0.131326 &
662{,}403 & 0.001148 \\
iTransformer \cite{liu2024itransformer} &
3.305829 & 2.614187 & 10.928505 &
0.075758 & 0.038648 & 0.105390 &
814{,}273 & 0.001347 \\
FEDformer \cite{zhou2022fedformer} &
0.598971 & 0.393850 & 0.358766 &
0.832540 & 0.855164 & 0.970631 &
756{,}569 & 0.001120 \\
Informer \cite{zhou2021informer} &
0.367830 & 0.193562 & 0.135299 &
0.897161 & 0.928818 & 0.988924 &
1{,}256{,}449 & 0.001365 \\
TFT \cite{lim2021temporal} &
0.421653 & 0.240557 & 0.177791 &
0.882113 & 0.911535 & 0.985446 &
2{,}510{,}702 & 0.001377 \\
ETSformer \cite{woo2022etsformer} &
0.333457 & 0.230507 & 0.111193 &
0.906772 & 0.915231 & 0.990898 &
800{,}376 & 0.001143 \\
Crossformer \cite{zhang2023crossformer} &
0.295324 & 0.175972 & 0.087842 &
0.913117 & 0.932184 & 0.991612 &
1{,}591{,}321 & 0.001349 \\
\bottomrule
\end{tabular}%
}
\\[2pt]
\raggedright\scriptsize
\textit{Notes:} Inference timing on GPU was measured using synchronized CUDA events~\cite{nvidia_cuda_events, pytorch_cuda_event}.
\end{table*}

\section{Performance and Numerical Results}
\label{sec:perf-numerical}

\subsection{Benchmark Results and Overall Comparison}
\label{subsec:results_interpretation}


\subsubsection{Accuracy---Control-grade Prediction with Competitive Error}
\label{subsubsec:acc}

As shown in Table~\ref{tab:results-liqss}, LiQSS attains Root Mean Squared Error (RMSE) = 0.2866, MAE = 0.1641, and MSE = 0.0821 on next-step RSRP forecasting.
Among the SSM family, LiQSS improves upon the strongest SSM baselines:
\begin{itemize}[leftmargin=1.2em]
    \item \textbf{vs.\ WM--MS$^{3}$M:} RMSE improves by $\approx 1.74\%$, Mean Absolute Error (MAE) by $\approx 1.99\%$, and MSE by $\approx 3.46\%$.
    \item \textbf{vs.\ MS$^{3}$M:} RMSE improves by $\approx 1.78\%$, MAE by $\approx 3.65\%$, and MSE by $\approx 3.54\%$.
\end{itemize}
This indicates that the proposed \emph{HiPPO--LegS mixture dynamics} plus \emph{quantum-inspired tensor-network parameterization} provides a measurable gain beyond prior structured-SSM forecasters, despite operating under the same strictly causal and leakage-safe evaluation pipeline.

Across Transformer-based methods, LiQSS remains strongly competitive and typically superior.
For example, compared to Chronos-GPT, LiQSS reduces RMSE by $\approx 35.67\%$ and MAE by $\approx 10.37\%$.
Methods such as RWKV, Performers, RetNet, PatchTST, and iTransformer exhibit substantially larger errors in this setting,
suggesting that attention/Transformer capacity does not automatically translate to control-grade accuracy under
Near-RT constraints and this specific O-RAN telemetry regime.

Crossformer attains RMSE $=0.295324$ in Table~\ref{tab:results-liqss}, which is slightly higher than LiQSS (RMSE $=0.286599$), i.e., $\approx 3.05\%$ worse in RMSE. However, this comes with \emph{much higher} deployment cost; Crossformer is $\approx 36\times$ larger in parameters and $\approx 2.96\times$ slower in inference than LiQSS (see \Cref{subsubsec:efficiency}). Therefore, LiQSS provides the best \emph{accuracy--efficiency} trade-off and dominates the practical Near-RT operating point.

\subsubsection{Forecast Skill and Goodness-of-Fit}
\label{subsubsec:skill}

Beyond absolute error metrics (RMSE/MAE/MSE), Table~\ref{tab:results-liqss} reports
\emph{forecast skill scores} that quantify relative performance against
simple, deployment-relevant reference predictors.
Relative evaluation against such benchmarks is standard practice in the
forecasting literature, particularly when assessing practical predictive utility
beyond naive baselines \cite{Hyndman2006Foresight}.

Let $\mathcal{E}(\hat{y},y)$ denote the mean-squared error (MSE) over the evaluation set $
\mathcal{E}(\hat{y},y)
\;=\;
\frac{1}{|\mathcal{D}|}\sum_{t\in\mathcal{D}}
\bigl(\hat{y}_{t+1}-y_{t+1}\bigr)^2,
$ and let $\hat{y}^{\text{model}}_{t+1}$ be the one-step-ahead forecast produced by the model.

\paragraph{Skill(R)---Persistence Skill}
The persistence (random-walk) baseline assumes that the next value equals the most
recently observed value $\hat{y}^{(R)}_{t+1} = y_t $. The corresponding skill score is defined as, $
\mathrm{Skill}(R)
\;=\;
1-\frac{\mathcal{E}(\hat{y}^{\text{model}}_{t+1},y_{t+1})}
        {\mathcal{E}(\hat{y}^{(R)}_{t+1},y_{t+1})},
$ following the standard MSE-based skill-score formulation
\cite{Murphy1988Skill}.
Thus, $\mathrm{Skill}(R)=1$ indicates perfect forecasting,
$\mathrm{Skill}(R)=0$ indicates parity with persistence,
and $\mathrm{Skill}(R)<0$ indicates worse-than-persistence performance.
For short-horizon time-series forecasting, persistence is a strong and highly relevant
baseline \cite{Hyndman2006Foresight}.

\paragraph{Skill(M)---Mean Skill}
The mean baseline predicts a constant value equal to the training-set mean
$\hat{y}^{(M)}_{t+1} = \mu_{\mathrm{tr}}$. The associated skill score is $
\mathrm{Skill}(M)
\;=\;
1-\frac{\mathcal{E}(\hat{y}^{\text{model}}_{t+1},y_{t+1})}
        {\mathcal{E}(\hat{y}^{(M)}_{t+1},y_{t+1})},
$ which quantifies improvement over a stationary reference predictor using the same
skill-score normalization \cite{Murphy1988Skill}.

\paragraph{LiQSS Model}
Using MSE-based skill scores on the test set, LiQSS achieves
$\mathrm{Skill}(R)=0.9199$ and $\mathrm{Skill}(M)=0.9397$,
indicating that its test MSE is reduced to $8.01\%$ of the persistence baseline and
$6.03\%$ of the mean baseline, respectively.
Together with a high coefficient of determination ($R^2=0.9933$, test set),
these results demonstrate that LiQSS accurately captures both short-horizon dynamics
(beating the strong persistence reference) and the overall variance structure
(beating the mean reference), while preserving strict causality and linear-time execution.

\subsubsection{Efficiency---Parameter count, Latency, and Near-RT Suitability}
\label{subsubsec:efficiency}

The most deployment-relevant outcome is that LiQSS achieves high accuracy at a tiny cost of
\emph{44{,}109 parameters} and \textbf{\SI{0.000456}{\second}} inference time per window on the target GPU.
This yields:
\begin{itemize}[leftmargin=1.2em]
    \item \textbf{vs.\ WM--MS$^{3}$M:} $\approx 90.77\%$ fewer parameters (about $10.8\times$ smaller) and
    $\approx 1.42\times$ faster inference.
    \item \textbf{vs.\ MS$^{3}$M:} $\approx 93.69\%$ fewer parameters (about $15.8\times$ smaller) and
    $\approx 1.41\times$ faster inference.
    \item \textbf{vs.\ Crossformer:} $\approx 97.23\%$ fewer parameters (about $36\times$ smaller) and
    $\approx 2.96\times$ faster inference.
    \item \textbf{vs.\ Chronos-GPT:} $\approx 99.36\%$ fewer parameters (about $155\times$ smaller) and
    $\approx 1.64\times$ faster inference.
\end{itemize}

These gains directly support Near-RT RIC operation where model size, memory traffic, and tail latency often dominate
end-to-end xApp responsiveness.


\subsection{Qualitative KPI Prediction and Temporal Fidelity}

\begin{figure}[t!]
\centering
\includegraphics[width=0.95\linewidth]{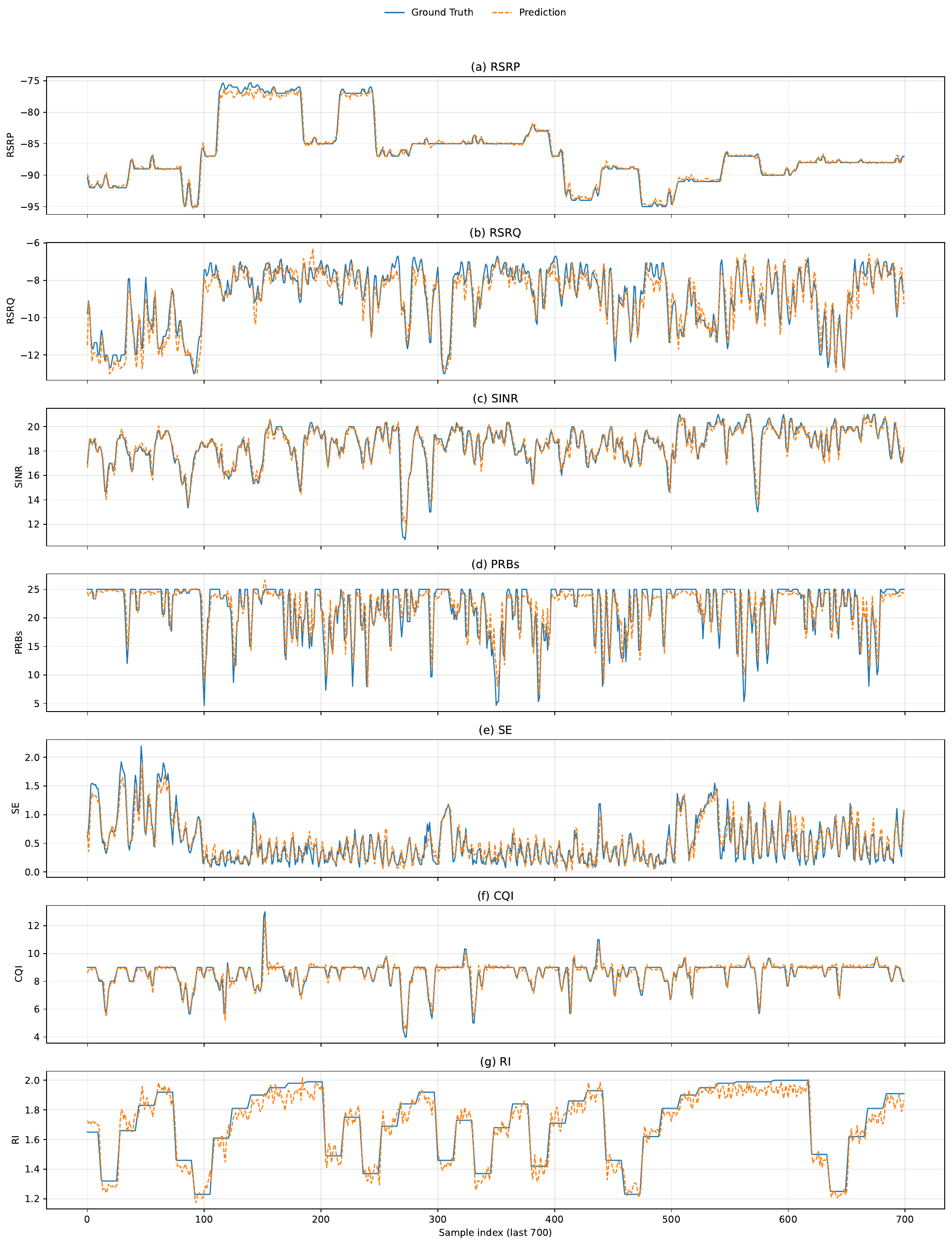}
\caption{Ground truth versus one-step-ahead predictions for key O-RAN KPIs over the last 700 test samples.}
\label{fig:gt-pred-kpis}
\end{figure}

Figure~\ref{fig:gt-pred-kpis} illustrates qualitative one-step-ahead forecasting performance across seven representative KPIs, comparing ground truth telemetry against the model’s predictions over the final 700 test samples.  Across radio-quality indicators (RSRP, RSRQ, and SINR), the model closely tracks both slow-varying trends and short-term fluctuations, with small residual errors except at abrupt regime transitions. This indicates that the state-space backbone captures dominant channel dynamics without relying on future information. For control-related and load-sensitive KPIs (PRBs and SE), predictions remain well aligned with ground truth, including sharp drops and bursts that reflect scheduler-driven behavior. This is particularly important because PRBs act as explicit control inputs during planning, and accurate reconstruction confirms internal consistency between action channels and predicted network responses. Discrete or quasi-discrete indicators (CQI and RI) exhibit piecewise-constant behavior with occasional jumps. The model successfully anticipates most transitions, while minor misalignments occur near abrupt state changes—an expected limitation under strictly causal, one-step forecasting. Overall, the figure confirms that the proposed LiQSS model delivers high-fidelity, leakage-safe predictions across heterogeneous KPIs, supporting accurate next-step forecasting for Near-RT O-RAN control.

\subsection{Empirical Evidence of Linear Scaling}

\begin{figure}[t!]
\centering
\includegraphics[width=0.8\linewidth]{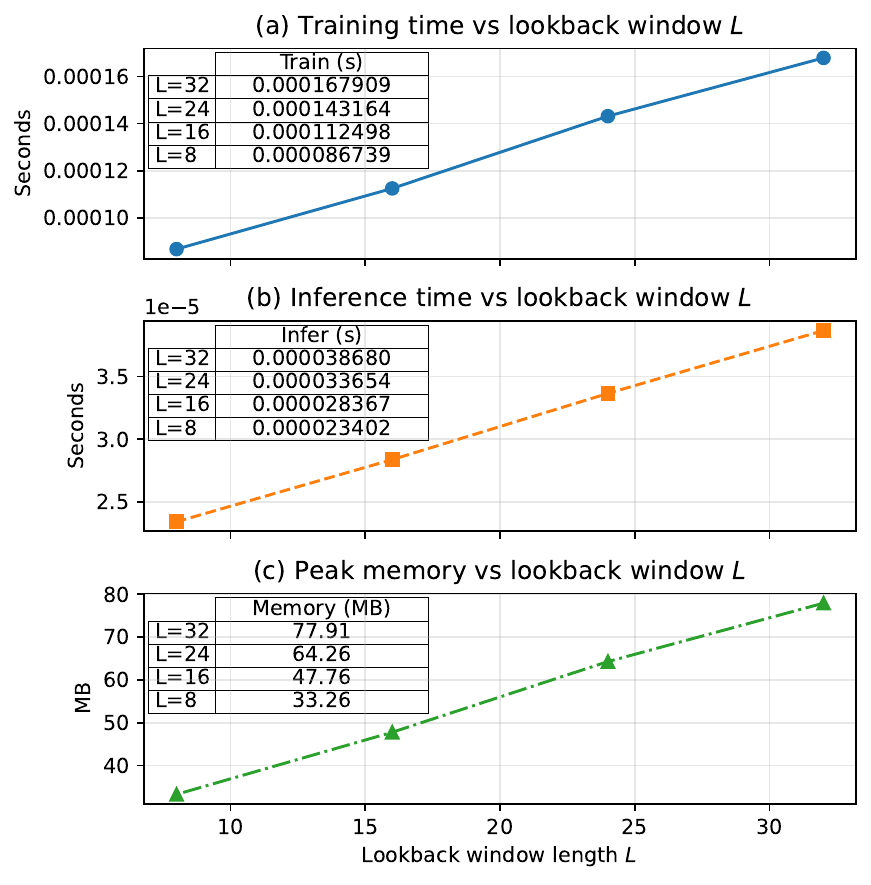}
\caption{Empirical evidence of linear-time behavior in LiQSS. Here, training time is the average seconds per example spent on one training step (forward pass + loss + backward pass + optimizer update) measured on a fixed batch after warm-up.}

\label{fig:liqss_window}
\end{figure}

Figure~\ref{fig:liqss_window} empirically demonstrates that LiQSS exhibits linear behavior in practice across training time, inference time, and memory usage as the lookback window length $L$ increases.
Figure~\ref{fig:liqss_window}-(a) shows that training time grows approximately proportionally with $L$, increasing smoothly from $8.67\times10^{-5}$~s at $L{=}8$ to $1.68\times10^{-4}$~s at $L{=}32$, confirming that backpropagation preserves linear scaling with respect to the lookback window length. Figure~\ref{fig:liqss_window}-(b) confirms the same trend at inference time, where latency increases linearly from $2.34\times10^{-5}$~s to $3.87\times10^{-5}$~s as $L$ quadruples, ruling out hidden quadratic costs during deployment.
Figure~\ref{fig:liqss_window}-(c) further shows that peak memory consumption scales linearly with $L$, growing from $33.26$~MB to $77.91$~MB, consistent with storing only $\mathcal{O}(L D)$ activations rather than $\mathcal{O}(L^2)$ attention maps.
Together, these subplots confirm that the architectural linearity of LiQSS directly translates into real-world execution.

\subsection{Sensitivity Analysis}
\label{subsec:sensitivity}

We analyze how LiQSS changes with three key capacity knobs while keeping the dataset, leakage-safe preprocessing, training protocol, and all other hyperparameters fixed: (i) the TT rank $r$ (controls coupling capacity in the TT/MPS input and head), (ii) the number of HiPPO--LegS mixture components per block $C_m$ (controls multi-timescale kernel capacity), and (iii) the SSM state dimension $N_s$ (controls temporal memory per channel). The results in Table~\ref{tab:sensitivity} show that performance is generally stable with modest and sometimes non-monotonic trends. Increasing TT rank yields only incremental accuracy gains (best RMSE at $r{=}16$) but increases latency, making $r{=}4$ the best Near-RT operating point. Increasing mixture components improves slightly up to $C_m{=}4$, but larger mixtures ($C_m{\ge}6$) degrade accuracy while substantially increasing parameters, indicating diminishing returns/over-parameterization for one-step forecasting. For the SSM state dimension, very small $N_s$ underfits, while overly large $N_s$ increases cost and can hurt accuracy; $N_s{=}32$ provides the best overall balance. Overall, the default configuration $(r{=}4,\,C_m{=}2,\,N_s{=}32)$ remains the strongest accuracy--efficiency choice for Near-RT deployment, with $(r{=}16,\,C_m{=}4,\,N_s{=}32)$ as an accuracy-leaning alternative when latency budget allows.

\begin{table}[t] \centering \footnotesize \setlength{\tabcolsep}{13pt} \caption{Sensitivity Test of LiQSS.} \label{tab:sensitivity} \resizebox{\columnwidth}{!}{\begin{tabular}{lrrrr} \toprule \textbf{Setting} & \textbf{RMSE} & \textbf{\#Params} & \textbf{Infer (s)} &  $\boldsymbol{R^2}$ \\ \midrule \multicolumn{5}{l}{\emph{\textbf{TT rank $r$ (with $C_m{=}2$, $N_s{=}32$)}}} \\ $r{=}2$ & 0.288711 & 43{,}885 & 0.000549 & 0.993177 \\ $r{=}4$ & 0.286599 & 44{,}109 & 0.000456 & 0.993276 \\ $r{=}8$ & 0.290643 & 44{,}749 & 0.000584 & 0.993085 \\ $r{=}16$ & 0.283614 & 46{,}797 & 0.000727 & 0.993415 \\ \midrule \multicolumn{5}{l}{\emph{\textbf{Mixture components $C_m$ (with $r{=}4$, $N_s{=}32$)}}} \\ $C_m{=}2$ & 0.286599 & 44{,}109 & 0.000456 & 0.993276 \\ $C_m{=}4$ & 0.284246 & 60{,}753 & 0.000593 & 0.993386 \\ $C_m{=}6$ & 0.286833 & 77{,}397 & 0.000656 & 0.993265 \\ $C_m{=}8$ & 0.291332 & 94{,}041 & 0.000623 & 0.993052 \\ \midrule \multicolumn{5}{l}{\emph{\textbf{SSM state dimension $N_s$ (with $r{=}4$, $C_m{=}2$)}}} \\ $N_s{=}8$ & 0.297041 & 31{,}821 & 0.000564 & 0.992777 \\ $N_s{=}16$ & 0.291094 & 35{,}917 & 0.000462 & 0.993064 \\ $N_s{=}32$ & 0.286599 & 44{,}109 & 0.000456 & 0.993276 \\ $N_s{=}64$ & 0.292463 & 60{,}493 & 0.000616 & 0.992998 \\ \bottomrule \end{tabular}} \end{table}

\section{Conclusion}
\label{sec:conclusion}

In this paper, we have proposed \emph{LiQSS}, a post-Transformer, quantum-inspired linear state-space tensor-network architecture for control-grade KPI forecasting in Near-RT O-RAN systems. By replacing quadratic self-attention with stable HiPPO--LegS state-space dynamics implemented via causal depthwise convolutions, and by constraining global linear operators through TT/MPS parameterization, LiQSS achieves strictly linear-time sequence modeling with a compact model footprint. Extensive evaluation on a bespoke O-RAN telemetry dataset comprising 59{,}441 sliding windows across 13 KPIs has demonstrated that LiQSS delivers forecasting accuracy superior to, state-of-the-art Transformer and structured-SSM baselines, while reducing trainable parameter count and significantly lowering inference latency. These results support the viability of post-Transformer, state-space, and tensor-network-based models as enabling technologies for deployment-faithful intelligence in future 6G Near-RT RIC environments. 

\bibliographystyle{IEEEtran}
\bibliography{refs}

\end{document}